\documentclass{article}

\newcommand{\diff}{\mathrm{d}}
\usepackage[aux]{rerunfilecheck}
\usepackage{amsmath}
\usepackage{amssymb}
\usepackage{mathtools}
\usepackage[autostyle]{csquotes}
\usepackage{xfrac}
\usepackage{float}
\usepackage{scrhack}
\usepackage{graphicx}
\usepackage{color}
\usepackage{bbm} 
\usepackage[section, below]{placeins} 
\usepackage[
    labelfont=bf,        
    font=small,          
    format=plain,
    width=0.9\textwidth, 
]{caption}
\usepackage{subcaption}
\usepackage{grffile}
\usepackage{booktabs}

\usepackage[
    sorting = none,
    sortcites = true,
    citestyle = numeric -comp, 
    backend = biber
]{biblatex}

\usepackage[
    unicode,        
    pdfusetitle,    
    pdfcreator={},  
    pdfproducer={}, 
]{hyperref}

\usepackage{bookmark}
\usepackage[shortcuts]{extdash}

\title{\bf Quantum Gravitational Decoherence in the 3 Neutrino Flavor Scheme}

\author{Dominik Hellmann$^1$, Heinrich P\"as$^1$, Erika Rani$^{1,2}$
\smallskip
\\
{\it $^1$ Fakult\"at f\"ur Physik,
Technische Universit\"at Dortmund,
Germany}
\\
{\it $^2$ UIN Maulana Malik Ibrahim Malang, Indonesia}
}

\begin{document}

\maketitle

\begin{abstract}
In many theories of quantum gravity quantum fluctuations of spacetime may serve as an environment for decoherence.
Here we study quantum-gravitational decoherence of high energy astrophysical neutrinos in the presence of fermionic dark sectors and for a realistic three neutrino scenario.
We show how violation of global symmetries expected to arise in quantum gravitational interactions provides a possibility to pin down the number of dark matter fermions in the universe.
Furthermore, we predict the expected total neutrino flux and flavor ratios at experiments depending on the flavor composition at the source.
\end{abstract}

\section{Introduction}
\label{sec:intro}
The search for a reliable and consistent theory of dark matter (DM) and the quest for a testable theory of quantum gravity (QG) are two of the most important open topics in modern physics.
Due to the feeble interaction of both sectors with known matter, it is difficult to find out if any of the existing theoretical models is realized in nature.
In the case of DM\footnote{Assuming that DM is comprised of new particle species.}, this is because it is most likely made up from singlets with respect to the unbroken Standard Model (SM) gauge group \(SU(3)_c \times U(1)_{\mathrm{EM}}\) and thus at most interact weakly (if at all) with our experimental set-ups.
For QG on the other hand, we know that all types of matter are fundamentally linked to spacetime and the dynamics of one also influences the dynamics of the other, but the energy scale at which quantum gravity effects would become visible, i.e. at the Planck scale, is far outside of the reach of today's experiments.
Hence, we need to employ an indirect mechanism to learn more about either of both subjects.\\
In this work, we analyze for the first time quantum gravitational decoherence in a complete system with three light neutrinos and \(n\) additional fermions from dark sectors.
This extends our previous discussions from~\cite{Hellmann:2021jyz} where we have proposed a possibility to search for such indirect effects from both new physics sectors by examining a single and possibly sensitive physical system, astrophysical neutrinos.\\
As mentioned above, all kinds of matter are indirectly coupled to each other due to their interaction with spacetime.
If spacetime itself has quantum properties it is subject to quantum fluctuations manifesting themselves for example as Planck scale black holes~\cite{Wheeler:1955zz,Amelino-Camelia:2008aez}.
Subsequently, every system of particles \(S\) evolving through spacetime is coupled to this dynamical environment \(\mathcal{E}\).
Interactions between degrees of freedom in \(S\) and \(\mathcal{E}\) cause quantum decoherence if only information about \(S\) is accessible~\cite{Zeh:1970zz}.
If, furthermore, virtual black holes in spacetime foam also obey the \textit{no-hair} theorem~\cite{Israel:1967wq,Hawking:1975vcx,Page:1980qm}, their interaction with propagating particles would violate global quantum numbers, such as lepton and flavor numbers.
In the following, we assume that this is the case and show how this property can be exploited by considering astrophysical neutrinos to pin down the number of neutral fermions in a certain mass range.\\
As many fermionic dark matter candidates~\cite{Dodelson:1993je,Jungman:1995df,SungCheon:2007nw,Kim:2008pp}, neutrinos are \(SU(3)_c \times U(1)_{\mathrm{EM}}\) singlets, too.
According to the assumed flavor blindness of QG interactions, a propagating neutrino system will therefore develop DM components after a sufficiently long distance.
This, subsequently, leads to different oscillation signatures in neutrino oscillation experiments carrying the imprint of the DM fermions.\\
Of course this effect might be damped if dark matter particles do not carry the same weak isospin quantum numbers as neutrinos, but since \(SU(2)_L \times U(1)_Y\) is broken interactions with the Higgs field would allow for transitions between particles of different weak isospin at the cost of damping by appropriate factors of \(\sfrac{m}{E}\) (mass over energy of the particles).\\
One reason why no quantum gravitational decoherence effects have been observed yet~\cite{Fogli:2007tx,Coloma:2018idr} may be that these effects are very weak and are most likely to manifest themselves in systems which traveled a very long distance and carry very high energies.
Astrophysical neutrinos fulfill both requirements and hence might represent the most sensitive possibility to test these kinds of effects.\\
From now on, we consider a system of three mixed, active neutrino species and \(n\) additional neutral fermions using the density matrix formalism of open quantum systems.
The application of open quantum system techniques to mixed particle systems under the influence of QG interactions was pioneered by Ellis, Hagelin, Nanopoulos and Srednicki (EHNS) in the Kaon system~\cite{ELLIS1984381,Ellis:1992dz} (see also~\cite{Huet:1994kr}) and later applied to systems of two and three neutrino generations~\cite{Liu:1997zd,Chang:1998ea,Benatti:2000ph,Klapdor-Kleingrothaus:2000kdx,Gago:2002na,Barenboim:2006xt}.
To extend this approach to an arbitrary number of fermions, we consider the time evolution of the density matrix \(\rho(t)\) defined on the \(n + 3\) dimensional Hilbert space \(\mathcal{H}\) of flavor configurations.
In general, the time evolution of the state of the full system \(S + \mathcal{E}\) is governed by a Hamiltonian \(H_{S + \mathcal{E}}\).
In case this Hamiltonian is partially unknown or too complicated, one can resort to taking the partial trace over the degrees of freedom of \(\mathcal{E}\), i.e.
\begin{align}
    \frac{\diff}{\diff t}\rho_{S+\mathcal{E}}(t) &= -i[H_{S+\mathcal{E}}, \rho_{S+\mathcal{E}}(t)] \nonumber\\
    \stackrel{\mathrm{Tr}_{\mathcal{E}}}{\longrightarrow} \quad
    \frac{\diff}{\diff t}\rho_{S}(t) &= -i[H_{S}, \rho_{S}(t)] + \mathcal{D}[\rho_{S}(t)]\,,\label{eq:lindblad}
\end{align}
yielding the so called Lindblad equation\footnote{In order for this equation to hold, we need to assume that \(S\) and \(\mathcal{E}\) are only weakly coupled which is a reasonable assumption in our case since we are considering quantum gravity effects on a beam of particles travelling through spacetime.} of the system \(S\).
The \textit{dissipator} \(\mathcal{D}\) arising in the process is identically zero if \(S\) and \(\mathcal{E}\) do not interact, i.e. \(H_{S + \mathcal{E}} = I_{\mathcal{E}} \otimes H_{S} + H_{\mathcal{E}} \otimes I_{S}\), but is non-zero if a term \(H_{\mathrm{int}}\) exists coupling \(S\) and \(\mathcal{E}\).
Hence, it describes the effective influence of \(\mathcal{E}\) on \(S\) while \(H_{S}\) only incorporates the physics of \(S\) itself detached from the environment.\\
In the full description of the system, this coupling gives rise to entanglement between degrees of freedom in \(S\) and \(\mathcal{E}\), but in our approximation it gives rise to mixed states \(\mathrm{Tr}(\rho_S^2) < 1\), i.e. decoherence.
This is just the consequence of the fact, that our system is not properly described by the degrees of freedom in \(S\) and hence we can only give probabilities in which quantum state our system is in.\\
In the following sections, we discuss how such a dissipator can be modeled for quantum gravity effects influencing our \(n + 3\) level flavor system.

\section{Modelling Quantum Gravity Effects}
\label{sec:model}
For the light Standard Model neutrino mass eigenstates \(\nu_k\) propagating in vacuum, we employ the usual ultra relativistic approximation
\begin{align}
    E_k = \sqrt{p^2 + m_k^2} \approx p + \frac{m_k^2}{2p}\,.
\end{align}
Note that this approximation is used for convenience but is not necessary in order to derive the results in the following.
Subsequently the 3 neutrino Hamiltonian reads
\begin{align}
    H_{S} &= p I_S + \frac{1}{2p}\mathrm{diag}(m_1^2, m_2^2, m_3^2)\,.
\end{align}
Since only the commutator of \(H_S\) and \(\rho\) impacts the evolution of the system, we can always subtract a part proportional to the identity \(I_S\) from the Hamiltonian.
Hence, it simplifies to
\begin{align}
    H_{S} &= \frac{1}{2p} \mathrm{diag}(0 , \Delta m_{21}^2, \Delta m_{31}^2)\,,
\end{align}
with \(\Delta m_{ji}^2 := m_j^2 - m_i^2\).
Furthermore, we include \(n\) additional fermions either carrying the same gauge quantum numbers as neutrinos or being Standard Model (SM) gauge singlets.
In order to obtain a similar Hamiltonian for this generalized case, we need to require \(p \gg m_{\mathrm{max}} := \max(\{m_k\}_{k = 1}^{n + 3})\) where \(m_1, m_2, m_3\) are the neutrino masses and the remaining ones correspond to the additional fermions.
Therefore, the Hamiltonian of the full system reads
\begin{align}
    H_{S} &= \frac{1}{2p} \mathrm{diag}(\{\Delta m_{j1}^2\}_{j = 1}^{n + 3})\,.
\end{align}
Until now we only discussed the details of the coherent evolution of the system and it is time to turn towards the modelling of the decoherence effects.
In total there are two important effects which have to be taken into account:
\begin{enumerate}
    \item Wave packet (WP) separation
    \item Quantum gravitational (QG) induced decoherence
\end{enumerate}
The first one arises because realistic neutrino states always occur as a superposition of finitely sized wave packets with spatial width \(\sigma_x\) since they are produced in processes of finite duration.
These wave packets of different mass eigenstates don't travel at the same group velocity \(v_k = \sfrac{\diff E_k}{\diff p}\) due to the different masses of the \(\nu_k\).
Thus, after some coherence length \(L_{jk}^{\mathrm{wp}}\) the wave functions of \(\nu_j\) and \(\nu_k\) barely overlap and coherence is lost\footnote{If the measurement process occurs in sufficiently short time before the other mass eigenstate wave packets arrive at the detector, it can distinguish between the different states. Only in this case the coherence is lost.}.
In the simplest meaningful model this can be described by exponential damping of the off-diagonal elements of the density matrix~\cite{Akhmedov:2009rb, Akhmedov:2014ssa} in the mass basis, i.e.
\begin{align}
    \mathcal{D}_{\mathrm{wp}} &= -\sum_{j > i} \frac{1}{L_{ji}^{\mathrm{wp}}} \mathbb{T}_{ji}\,,
\end{align}
where \(\mathbb{T}_{ji}\) applied to the density matrix yields the same matrix with all elements set to zero but the entries \(\rho_{ji}\) and \(\rho_{ij}\) and the coherence length is given by
\begin{align}
    L_{ji}^{\mathrm{wp}} &= \frac{\sigma_x}{\vert\Delta v_{ji}\vert} \approx \sigma_x \frac{2 p^2}{\vert\Delta m_{ji}^2\vert}\,. \label{eq:wp_param}
\end{align}
Quantum gravitationally induced decoherence on the other hand is due to the interaction of the system with the spacetime foam.
Following EHNS, we assume that in each of these interactions the \textit{no-hair theorem} applies and all information about the flavor composition of the state is lost.
Hence, considering an ensemble of initially pure flavor states encountering these stochastic spacetime interactions, we find that after a sufficiently long travel distance the system gets maximally entropic since no information of the initial flavor can be restored.
This corresponds to an uniform flavor distribution.

\subsection{A Useful Set of Basis Matrices}
Before discussing the form of the QG dissipator, we introduce a useful set of basis matrices in which we will expand the density matrix.
This is the set of hermitian \(SU(N := n + 3)\) generators (plus a matrix proportional to the identity) \(\{\lambda\}_{k = 0}^{N^2 - 1}\) where we already adjusted the dimension of the group to fit our \(n + 3\) level system.
These basis matrices fulfill the following criteria:
\begin{itemize}
    \item Orthonormality: \\
        \(\langle \lambda_j, \lambda_k \rangle := 2\mathrm{Tr}(\lambda_j \cdot \lambda_k) = \delta_{jk}\)
    \item Trace Identities: \\
        \(\mathrm{Tr}(\lambda_j) = 0\) iff \(j = 1, \ldots, N^2 - 1\) and \(\mathrm{Tr}(\lambda_0) = \sqrt{\sfrac{N}{2}}\)
    \item Commutation relations: \\
        \([\lambda_0, \lambda_j] = 0\), \(\forall j = 0, \ldots, N\) and \([\lambda_j, \lambda_k] = i \sum_{l = 1}^{N^2 - 1} f_{jkl} \lambda_l\)
\end{itemize}
where \(f_{jkl}\) are the totally antisymmetric \(SU(N)\) structure constants.\\
For practical reasons, we use the following ordering for the basis matrices:
\begin{align}
    \{\lambda_k\}_{k = 0}^{N^2 - 1} &= \{\lambda_0, \underbrace{\lambda_{1}, \ldots, \lambda_{N(N - 1)}}_{\mathrm{off-diagonal}}, \underbrace{\lambda_{N(N - 1) + 1}, \ldots, \lambda_{N^2 - 1}}_{\mathrm{diagonal}}\}\,,
\end{align}
with
\begin{align}
    (\lambda_{j})_{kl} &= \frac{1}{2}(a \delta_{kk_0}\delta_{ll_0} + a^{\ast} \delta_{lk_0}\delta_{kl_0}) \,, \quad 1 \leq j \leq N(N-1)\\
    \lambda_{N (N - 1) + m} &= \frac{1}{\sqrt{2 m(m + 1)}} \mathrm{diag}(\underbrace{1, \ldots, 1}_{m \times}, -m, 0, \ldots, 0) \,, \;\; 1 \leq m \leq N - 1\,.
\end{align}
For the off-diagonal matrices, we use an ordering such that \(a\) alternates between \(1\) and \(i\) for increasing index \(j\) and the indices \(k_0\) and \(l_0\) are arranged that
\begin{align}
    (k_0, l_0) = (2,1), (3, 1), \ldots, (N, 1), (3, 2), \ldots, (N, 2), \ldots, (N, N-1)\,,
\end{align}
where each tuple is attained twice: Once for \(a = 1\) and once for \(a = i\).
For example for \(N = 2\) we get the rescaled Pauli matrices and for \(N = 3\) we get a rearranged set of rescaled Gell-Mann matrices.
This rearrangement of the basis matrices implies that also the ordering of vector and matrix components is different from the usual ordering in the literature concerned with three neutrino oscillations with decoherence.

\subsection{The Lindblad Equation in the New Basis}
Using this basis, the Lindblad equation~\eqref{eq:lindblad} becomes
\begin{align}
    \frac{\diff \vec{\varrho}(x)}{\diff x} &= C \vec{\varrho}(x) + D \vec{\varrho}(x) := \Lambda \vec{\varrho}(x) \,, \label{eq:EoM}
\end{align}
where \(\vec{\varrho}\) is the coefficient vector of \(\rho\), \(C\) is the representation matrix of the commutator \(-i[H,\cdot]\) and \(D\) is the representation matrix of the dissipator \(\mathcal{D}\) in our basis.
Furthermore, we employ the ultra relativistic approximation in order to substitute the traveled path \(x\) for the time variable \(t\).\\
The antisymmetric commutator matrix \(C\) is given by
\begin{align}
    C_{kl} &= -\sum_{j = 1}^{N^2 - 1} h_{j} f_{jkl} = - C_{lk} \,, &\forall k,l = 1, \ldots, N^2 - 1 \,,\\
    C_{0l} &= 0 = - C_{l0} \,, &\forall l = 0, \ldots, N^2 - 1 \,,
\end{align}
where \(h_{l}\) is the coefficient vector of the Hamiltonian.\\
The action of \(\mathcal{D}^{\mathrm{wp}}\) on a given density matrix \(\rho\) is simple and only amounts to multiplying its off-diagonal elements by the appropriate negative inverse wave packet coherence lengths, \(-\sfrac{1}{L_{ji}^{\mathrm{wp}}}\).
Expressed in the chosen basis this corresponds to a diagonal dissipator of the form
\begin{align}
    D^{\mathrm{wp}} &= -\mathrm{diag}\left(0, \frac{1}{L_{21}^{\mathrm{wp}}}, \frac{1}{L_{21}^{\mathrm{wp}}},
    \frac{1}{L_{31}^{\mathrm{wp}}}, \frac{1}{L_{31}^{\mathrm{wp}}},
    \ldots, \frac{1}{L_{N\, N-1}^{\mathrm{wp}}}, \frac{1}{L_{N\, N-1}^{\mathrm{wp}}}, 0, \ldots, 0\right)\,.
\end{align}
Now, we return to the discussion of the dissipator matrix \(D^{\mathrm{qg}}\) corresponding to the quantum gravity effects.
Since we have no accepted theory of quantum gravity yet, we need to employ some basic assumptions in order to constrain the shape of \(D^{\mathrm{qg}}\).
At first, we assume that the effect is homogeneous and isotropic since there should be no preferred location or direction in the vacuum.
Therefore, \(D^{\mathrm{qg}}\) depends only on the average energy \(E\) of the system.
From now on, we use the approximation \(p \approx E\) for all formulas to align with the literature.
Second, we assume a universal power-law energy dependence~\cite{Lisi:2000zt,Anchordoqui:2005gj,Stuttard:2020qfv} of
\begin{align}
    D_{jk}^{\mathrm{qg}}(E) &= d_{jk} \frac{E^{\alpha}}{M_{\mathrm{Planck}}^{\alpha - 1}}\,, \label{eq:qg_param}
\end{align}
where \(\alpha\) and \(d_{jk}\) are free, dimensionless parameters of the model and \(M_{\mathrm{Planck}}\) is the Planck mass serving as the energy scale of the problem.\\
Next, we need to specify the shape of parameter matrix \(d\).
The requirement of monotonically increasing entropy, i.e. \(\sfrac{\diff S}{\diff t} \geq 0\), and probability conservation\footnote{We assume that our system does not loose particles, but only information.} \(1 \equiv \mathrm{Tr}(\rho) \propto \varrho_0\) yields
\begin{align}
    d_{0j} = d_{j0} = 0 \,, \quad \forall j = 0, \ldots, N^2 - 1\,.
\end{align}
Hence, we only need to consider the \((N^2 - 1) \times (N^2 - 1)\) submatrix \(\{d_{ij}\}_{i,j = 1}^{N^2 - 1}\).
In the following, we assume a symmetric dissipator because each matrix can be written as the sum of a symmetric and an antisymmetric matrix.
The antisymmetric matrix can then be directly compared to the commutator part whose entries are assumed to be much larger and hence we can neglect the effects of the antisymmetric part.
Therefore only the symmetric part of \(D\) gives rise to new, significant effects.\\
The simplest scenario fulfilling these criteria is a diagonal dissipator
\begin{align}
    d = \mathrm{diag}(0, d_{1}, \ldots, d_{N^2 - 1})\,.
\end{align}
Similarly to \(D^{\mathrm{wp}}\), this matrix results in a damping of all off-diagonal elements of \(\varrho\) at different rates but with the difference that \(d\) also damps the excess / lack of each particle species over flavor equilibrium according to the \(d_{j} \leq 0\) with \(j > N(N-1)\).
For this case, we can analytically solve the Lindblad equation.
The corresponding result will be shown in the next section.\\
The same asymptotic effect of convergence towards flavor equilibrium is achieved by all dissipators that have only one zero eigenvalue corresponding to the invariance of the trace with respect to time evolution\footnote{See App.~\ref{app:flav_eq} for details.}.
Hence this effect can be achieved by a much bigger class of dissipators than only diagonal ones, but since we are solely interested in the asymptotic limit in the following it is much simpler to resort to a diagonal \(D\).

\section{Oscillation Probabilities and Neutrino Fluxes}
\label{sec:probs}
Since in our approach the matrix \(\Lambda\) is independent of the traveled distance \(x\), the analytic solution of Eq.\eqref{eq:EoM} is given by
\begin{align}
	\vec{\varrho}(x) &= \exp(\Lambda (x - x_0)) \cdot \vec{\varrho}(x_0)\,.
\end{align}
Hence, our only remaining task is to choose a suitable \(\vec{\varrho}(x_0)\) for the scenarios we want to consider.
In all of these scenarios, we start with an ensemble of initially pure neutrino flavor eigenstates produced in an astrophysical environment via the weak interaction.
The simplest way to find the corresponding initial \(\vec{\varrho}(x_0)\), is to start from \(\vec{\varrho}^{f}(x_0)\) in the flavor basis and then transform it into the mass basis using the Pontecorvo Maki Nakagawa Sakata (PMNS) matrix \(U_{\mathrm{PMNS}}\).
Since we are not only considering the simple 3 neutrino case but also including \(n\) additional fermions into the system, we need to extend this mixing matrix as follows:
\begin{align}
	U &= U_{\mathrm{PMNS}} \oplus I_{n \times n} \,. \label{eq:mix}
\end{align}
Because neutrinos do not mix with the other fermions in the system and our initial state is a pure neutrino flavor eigenstate, it is sufficient to extend \(U_{\mathrm{PMNS}}\) using the identity, even though the other fermion species might also mix with each other.
Therefore, the transformation due to the matrix shown in~\eqref{eq:mix} is only a partial transformation to the neutrino flavor basis.
The initial density matrix in the mass basis is then given by
\begin{align}
	\rho(x_0) = \rho_{\alpha} &= U^{\dagger} \rho_{\alpha}^{f} U\,,
\end{align}
where \((\rho_{\alpha}^{f})_{ab} = \delta_{a \alpha} \delta_{b \alpha}\) is the neutrino flavor projector for flavor \(\nu_{\alpha}\).
The respective coefficient vector reads
\begin{align*}
	\varrho_{\alpha}^{0} &= \sqrt{\frac{2}{N}}\\
	\varrho_{\alpha}^{1} &= 2 \mathrm{Re}(U_{\alpha 2}^{\ast} U_{\alpha 1})\,,\\
	\varrho_{\alpha}^{2} &= 2 \mathrm{Im}(U_{\alpha 2}^{\ast} U_{\alpha 1})\,,\\
	\varrho_{\alpha}^{3} &= 2 \mathrm{Re}(U_{\alpha 3}^{\ast} U_{\alpha 1})\,,\\
	\varrho_{\alpha}^{4} &= 2 \mathrm{Im}(U_{\alpha 3}^{\ast} U_{\alpha 1})\,,\\
	\varrho_{\alpha}^{2N-1} &= 2 \mathrm{Re}(U_{\alpha 3}^{\ast} U_{\alpha 2})\,,\\
	\varrho_{\alpha}^{2N} &= 2 \mathrm{Im}(U_{\alpha 3}^{\ast} U_{\alpha 2})\,,\\
	\varrho_{\alpha}^{N(N-1)+1} &= \vert U_{\alpha 1} \vert^2 - \vert U_{\alpha 2} \vert^2\,,\\
	\varrho_{\alpha}^{N(N-1)+2} &= \frac{1}{\sqrt{3}}(\vert U_{\alpha 1} \vert^2 + \vert U_{\alpha 2} \vert^2 - 2\vert U_{\alpha 3} \vert^2)\,,\\
	\varrho_{\alpha}^{N(N-1)+k} &= \sqrt{\frac{2}{k(k+1)}}\,, \quad \forall 3 \leq k \leq N - 1 \,,
\end{align*}
all other components vanish.\\
Using these initial density matrices, we can calculate the oscillation probabilities as
\begin{align}
	P_{\alpha \beta}(L) &= \mathrm{Tr}(\rho_{\beta} \rho(L))\,, \quad \mathrm{with} \; \rho(0) = \rho_{\alpha}\\
	&= \frac{1}{2} \langle \rho_{\beta}, \rho(L) \rangle\\
	&= \frac{1}{2} \vec{\varrho}_{\beta}^{\,T} \vec{\varrho}(L)\\
	&= \frac{1}{2} \vec{\varrho}_{\beta}^{\,T} \exp(\Lambda L) \vec{\varrho}_{\alpha}\,.
\end{align}
For the simplest case of a purely diagonal dissipator, the general oscillation formula reads
\begin{align}
	P_{\alpha\beta}(L) = &\frac{1}{N} + \frac{1}{2} (\vert U_{\alpha 1} \vert^2 - \vert U_{\alpha 2} \vert^2) (\vert U_{\beta 1} \vert^2 - \vert U_{\beta 2} \vert^2) e^{-\Gamma_{N(N-1)+1}L}\nonumber\\
	&+ \frac{1}{6}(\vert U_{\alpha 1} \vert^2 + \vert U_{\alpha 2} \vert^2 - 2\vert U_{\alpha 3} \vert^2)(\vert U_{\beta 1} \vert^2 + \vert U_{\beta 2} \vert^2 - 2\vert U_{\beta 3} \vert^2) e^{-\Gamma_{N(N-1)+2}L} \nonumber\\
	&+ \sum\limits_{k = 3}^{N - 1} \frac{e^{-\Gamma_{N(N-1)+k}L}}{k(k+1)} \nonumber\\
	&+ 2 \sum\limits_{j > i = 1}^{3} \mathrm{Re}(U_{\alpha j}^{\ast}U_{\alpha i}U_{\beta j}U_{\beta i}^{\ast})e^{-\frac{L}{L_{ij}^{\mathrm{wp}}}}e^{-\bar{\Gamma}_{l+1 \, l}L}\cos(\omega_{ij}L)\nonumber\\
	&+ 2 \sum\limits_{j > i = 1}^{3} \mathrm{Re}(U_{\alpha j}^{\ast}U_{\alpha i}U_{\beta j}^{\ast}U_{\beta i}) \frac{\Delta \Gamma_{l+1 \, l}}{\omega_{ij}}e^{-\frac{L}{L_{ij}^{\mathrm{wp}}}}e^{-\bar{\Gamma}_{l+1 \, l}L}\sin(\omega_{ij}L) \nonumber\\
	&- 2 \sum\limits_{j > i = 1}^{3} \mathrm{Im}(U_{\alpha j}^{\ast}U_{\alpha i}U_{\beta j}U_{\beta i}^{\ast}) \frac{\Delta E_{ij}}{\omega_{ij}}
	e^{-\frac{L}{L_{ij}^{\mathrm{wp}}}}e^{-\bar{\Gamma}_{l+1 \, l}L}\sin(\omega_{ij}L)\,, \label{eq:prob}
\end{align}
where we introduce the following quantities
\begin{align}
	\Delta E_{ij} &:= \frac{\Delta m_{ij}^2}{2E}\,,\\
	\omega_{ij} &:= \sqrt{(\Delta E_{ij})^2 - (\Delta \Gamma_{l+1 \, l})^2}\,,\\
	\Delta \Gamma_{l + 1 \, l} &:= \frac{\Gamma_{l+1} - \Gamma_{l}}{2}\,,\\
	\bar{\Gamma}_{l+1 \, l} &:= \frac{\Gamma_{l+1} + \Gamma_{l}}{2}\,,\\
	\Gamma_{l} &:= \vert d_l \vert \frac{E^{\alpha}}{M_{\mathrm{Planck}}^{\alpha - 1}} \,,
\end{align}
and the index \(l\) is a function of the indices \(i\) and \(j\) such that
\begin{align}
	l(i,j) := \begin{cases} 1 & ,\,i = 1 \land j = 2 \\ 3 & ,\,i = 1 \land j = 3 \\ 2N -1 & ,\,i = 2 \land j = 3 \end{cases}\,.
\end{align}
For the more complicated scenarios where the QG dissipator also contains off-diagonal elements the solution has to be calculated semi-analytically.

\subsection{Asymptotic Limits}
\label{eq:asymp}
\begin{figure}
	\centering
	\includegraphics[width = 0.6\textwidth]{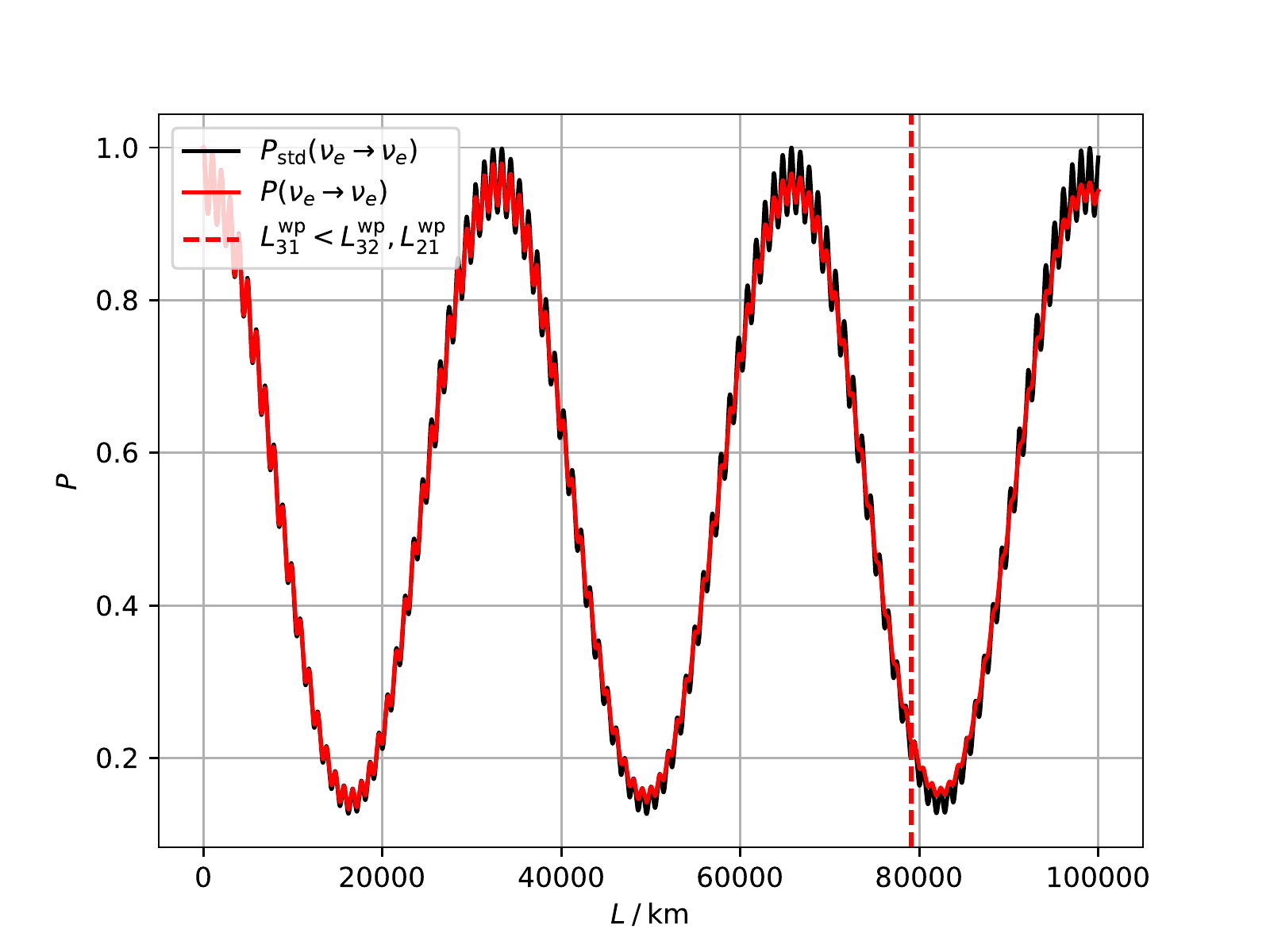}
	\caption{Comparison of the standard oscillation formula (black) versus the formula from Eq.~\eqref{eq:prob} (red) for \(p = 1\,\mathrm{GeV}\).}
	\label{fig:p_vs_std}
\end{figure}

\begin{figure}
    \begin{subfigure}[b]{0.45\textwidth}
		\centering
 		\includegraphics[width = \textwidth]{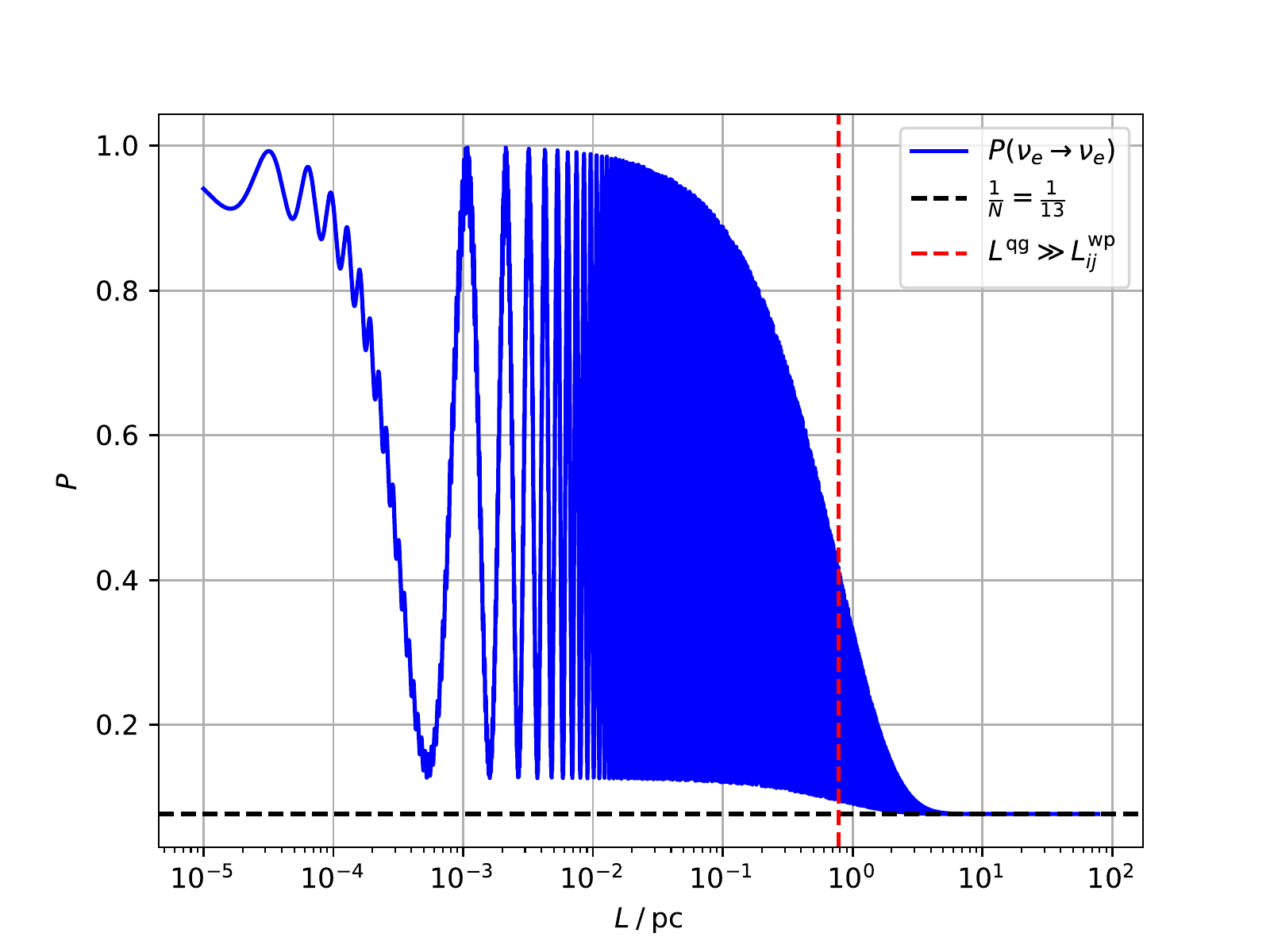}
 		\caption{}
 		\label{fig:p_asymp1}
    \end{subfigure}
    \hfill
    \begin{subfigure}[b]{0.45\textwidth}
		\centering
  		\includegraphics[width = \textwidth]{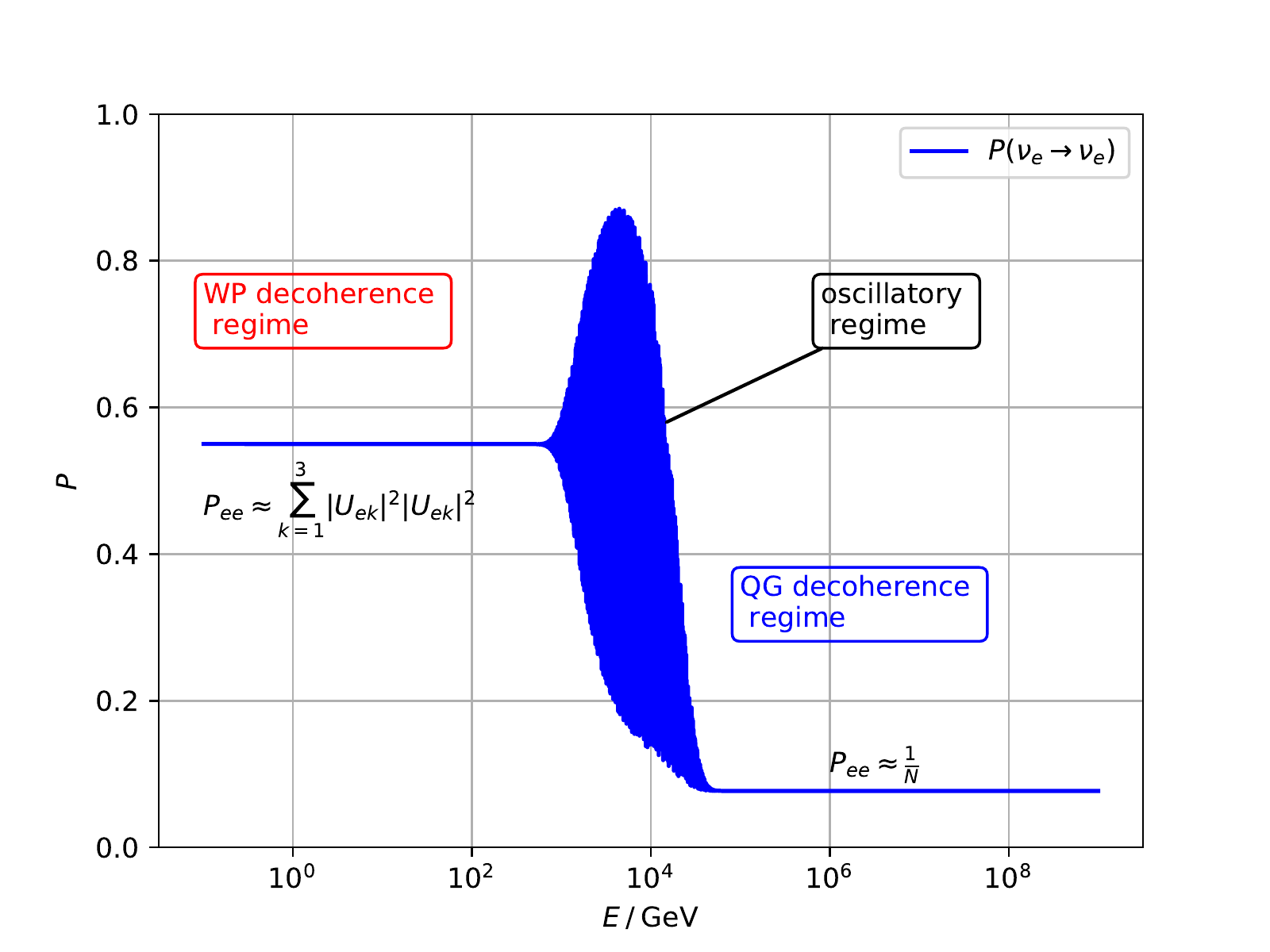}
  		\caption{}
  		\label{fig:p_asymp2}
	\end{subfigure}
    \caption{Oscillation probability \(P(\nu_e \rightarrow \nu_e)\) for variable base length (at \(E = 1\,\mathrm{PeV} \)) (left) and variable energy (at \(L = 2\,\mathrm{kpc}\)) (right).
	The \(L\) and \(E\) regions are chosen such that the asymptotic behavior of Eq.~\eqref{eq:prob} becomes appearant.
	Furthermore, we use \(\sigma_x = 10^{-9}\,\mathrm{m}\) in order to allow for an oscillatory regime in the right plot.}
    \label{fig:p_asymp}
\end{figure}

Now, we want to inspect the behavior of the formula just derived for some baselines of interest.
In the small baseline regime, where by small we mean small compared to all cohernce lengths \(L_{ij}^{\mathrm{wp}} \ll \Gamma_{k}^{-1}\) in the system, the standard 3 neutrino ocillation formula is recovered, \textit{i.e.}
\begin{align}
	P_{\alpha\beta}(L) \approx &\delta_{\alpha\beta} - 4 \sum\limits_{j > i} \mathrm{Re}(U_{\alpha j}^{\ast}U_{\alpha i}U_{\beta j}U_{\beta i}^{\ast}) \sin^2\left(\frac{\Delta m_{ji}^2 L}{4 E}\right)\nonumber\\
	&+ 2 \sum\limits_{j > i} \mathrm{Im}(U_{\alpha j}^{\ast}U_{\alpha i}U_{\beta j}U_{\beta i}^{\ast})  \sin\left(\frac{\Delta m_{ji}^2 L}{2 E}\right)\,, \label{eq:std}
\end{align}
but only if \(\Delta \Gamma_{l+1 \, l} \ll \Delta E_{ij}\) for the corresponding \(l(i,j)\).
This must be the case, otherwise we would already see significant discrepancies between the observed and predicted oscillation patterns in earth bound or solar neutrino oscillation experiments.
For a set of exemplary parameters given in Tab.~\ref{tab:params}, we plot the oscillation probability from Eq.~\eqref{eq:prob} against the standard probability~\eqref{eq:std} for baselines up to \(L \leq 10^{5}\,\mathrm{km}\) in Fig.~\ref{fig:p_vs_std}.
The plot shows the expected agreement of both curves for small baselines \(L \lesssim 2 \times 10^{4}\,\mathrm{km}\) and a growing difference between them as \(L\) approaches the smallest coherence length \(L_{31}^{\mathrm{wp}}\).\\
\begin{table}
	\centering
	\caption{Exemplary Parameter configuration used for the oscillation plots assuming normal ordering (NO) of the neutrino masses.}
	\label{tab:params}
	\begin{tabular}{ccc}
		\toprule
		Parameter & Value & Source \\
		\midrule
		\(\Delta m_{21}^2\) & \(7.53 \times 10^{-5}\,\mathrm{eV}^2\) & \cite{pdg} \\
		\(\Delta m_{32}^2\) (NO) & \(2.453 \times 10^{-3}\,\mathrm{eV}^2\) & \cite{pdg} \\
		\(\sin^2(\theta_{12})\) & \(0.307\) & \cite{pdg} \\
		\(\sin^2(\theta_{13})\) & \(2.18 \times 10^{-2}\) & \cite{pdg} \\
		\(\sin^2(\theta_{23})\) (NO) & \(0.545\) & \cite{pdg} \\
		\(\sigma_x\) & \(10^{-13}\,\mathrm{m}\) & \cite{Kersten:2015kio} \\
		\(N\) & 13 & - \\
		\(\alpha\) & \(2\) & - \\
		\(d := d_k\, \forall k \geq 1\) & \(-10^{-25}\) & - \\
		\bottomrule
	\end{tabular}
\end{table}
For baselines comparable to the coherence length induced by the effect of wave packet separation, the oscillation formula becomes
\begin{align}
	P_{\alpha\beta}(L) \approx &\sum\limits_{k = 1}^{3} \vert U_{\alpha k} \vert^2 \vert U_{\beta k} \vert^2 \nonumber\\
	&+ 2 \sum\limits_{j > i = 1}^{3} \mathrm{Re}(U_{\alpha j}^{\ast}U_{\alpha i}U_{\beta j}U_{\beta i}^{\ast})e^{-\frac{L}{L_{ij}^{\mathrm{wp}}}}\cos(\Delta E_{ij} L)\nonumber\\
	&- 2 \sum\limits_{j > i = 1}^{3} \mathrm{Im}(U_{\alpha j}^{\ast}U_{\alpha i}U_{\beta j}U_{\beta i}^{\ast})
	e^{-\frac{L}{L_{ij}^{\mathrm{wp}}}} \sin(\Delta E_{ij} L)\,.
\end{align}
Here, we assumed \(\Gamma_{k}^{-1} \gg L_{ij}^{\mathrm{wp}}\) and hence \(\exp(-\Gamma_k L) \approx 1\) which is reasonable since quantum gravity effects are supposed to be very weak.\\
For \( L \sim \Gamma_{k}^{-1}\), the asymptotic oscillation probability reads
\begin{align}
	P_{\alpha\beta}(L) \approx &\frac{1}{N} + \frac{1}{2} (\vert U_{\alpha 1} \vert^2 - \vert U_{\alpha 2} \vert^2) (\vert U_{\beta 1} \vert^2 - \vert U_{\beta 2} \vert^2) e^{-\Gamma_{N(N-1)+1}L}\nonumber\\
	&+ \frac{1}{6}(\vert U_{\alpha 1} \vert^2 + \vert U_{\alpha 2} \vert^2 - 2\vert U_{\alpha 3} \vert^2)(\vert U_{\beta 1} \vert^2 + \vert U_{\beta 2} \vert^2 - 2\vert U_{\beta 3} \vert^2) e^{-\Gamma_{N(N-1)+2}L} \nonumber\\
	&+ \sum\limits_{k = 3}^{N - 1} \frac{e^{-\Gamma_{N(N-1)+k}L}}{k(k+1)} \,,
\end{align}
which approaches flavor equilibrium, \textit{i.e.}
\begin{align}
	P_{\alpha\beta}(L \gg \Gamma_{k}^{-1}) \sim \frac{1}{N}\,,
\end{align}
iff \(\Gamma_{N(N-1) + k} \neq 0\) for \(k \geq 1\).\\
In view of what follows in the next subsections, we should also consider the behavior of the oscillation probabilities for different energy regimes.
Here, we note that wave packet separation is a low energy effect since
\begin{align}
	L_{ij}^{-1} \propto E^{-2}\,,
\end{align}
while quantum gravitationally induced decoherence dominates at high energies because
\begin{align}
	\Gamma_k \propto E^{\alpha}\,,
\end{align}
with \(\alpha \geq 1\), typically.
Depending on the decoherence parameters of the system there might exist a region between the wave packet separation and quantum gravity regimes where oscillations are dominating.
In Fig.~\ref{fig:p_asymp1} and~\ref{fig:p_asymp2}, we show the asymptotic behavior of \(P_{ee}\) for variable base length and energy, respectively.
For the plot at fixed base length, we choose \(L_S \approx 2\,\mathrm{kpc}\) which corresponds to the approximate distance of earth to Cygnus OB2 representing a potentially interesting source of \(\bar{\nu}_e\) according to~\cite{Anchordoqui:2005gj}.

\FloatBarrier

\subsection{Neutrino Fluxes}
\label{ssec:flux}
Since quantum gravity effects are expected to be extremely weak, we need to investigate on neutrinos of high energy which already traveled a significant distance from their origin to our detectors.
Therefore, we have to improve on measuring astrophysical neutrinos originating from quasars and other stellar objects which are boosted relative to earth such that they can reach energies of \(\mathcal{O}(\mathrm{PeV})\) or even \(\mathcal{O}(\mathrm{EeV})\).
If we are able to identify a rich high energy neutrino source using future neutrino experiments, we could be able to measure the previously described effects and hence learn something about the quantum nature of spacetime and dark matter fermions.
These sources will reside at a fixed baseline \(L_{S}\) and provide neutrinos of different energy.
Thus, we need to study the impact of quantum gravitational decoherence on the neutrino energy flux spectra corresponding to the neutrino sources.
In the following, we demonstrate how to estimate neutrino fluxes using the previously calculated neutrino oscillation probabilities.\\
A realistic neutrino source can be one of two kinds
\begin{itemize}
	\item Primary source: Neutrinos directly emerge from the approximately point like source
	\item Secondary source: The source produces particles (pions, neutrons, \ldots) which decay into neutrinos on their path to earth
\end{itemize}
In the first case, the flux density\footnote{Here flux density means the number of particles per area, time and energy, \textit{i.e.} \(\sfrac{\partial^3 N}{\partial A \partial t \partial E}\).} of neutrino flavor \(\nu_{\alpha}\) reaching earth \(\Phi_{\alpha}^{\oplus}(E)\) is given by
\begin{align}
	\Phi_{\alpha}^{\oplus}(E) &= \sum_{\beta = e}^{\tau} P_{\beta \alpha}(L_{S}, E) \Phi_{\beta}^{S}(E)\,, \label{eq:flux_simp}
\end{align}
where \(\Phi_{\beta}^{S}(E)\) is the flux density of neutrinos of flavor \(\beta\) emerging at the source\footnote{Of course this flux density is scaled apropriately such that it represents the flux of particles at earth if no oscillation effects would occur.}, \(L_{S}\) is the physical distance to the source and \(E\) is the neutrino energy.\\
The second case is a little bit more complicated since one has to take into account that the primary particles do not decay instantaneously but may travel for significant distances due to a huge Lorentz boost relative to the lab frame.
Furthermore, the dynamics and kinematics of the decay process need to be considered in order to translate the spectrum of primary particles to the neutrino spectrum.
Using the law of total probability one can derive the flux of neutrinos arriving at earth to be
\begin{align}
	\Phi_{\alpha}^{\oplus}(E) = \sum_{\eta \in S}\sum_{\beta = e}^{\tau} \; \int\limits_{m_{\eta}}^{\infty}\int\limits_{0}^{L_{S}} \Phi_{\eta}^{S}(E_{\eta})
	\pi_{\eta \beta}(E, E_{\eta}) \frac{e^{\frac{-\ell}{v_{\eta}\tau_{\eta}}}}{v_{\eta}\tau_{\eta}} P_{\beta \alpha}(E, L_{S} - \ell) \,\diff\ell \diff E_{\eta}\,, \label{eq:flux_gen}
\end{align}
which is a generalized version of the corresponding expression given in~\cite{Anchordoqui:2005gj}.
In the following, we briefly discuss the physical meaning of this formula.
We start with particles \(\eta\) (\textit{e.g.} \(\pi^{\pm}, n, \ldots\)) emerging from the source \(S\) with energy \(E_{\eta}\) and the flux density \(\Phi_{\eta}^{S}(E_{\eta})\).
These primary particles then decay according to the exponential distribution with mean lifetime \(\tau_{\eta}\) after a distance \(\ell\) from the source and become a neutrino of flavor \(\beta\) and energy \(E\) with the probability \(\pi_{\eta\beta}(E, E_{\eta})\).
These neutrinos travel the remaining distance \(L_{S} - \ell\) to earth and are measured at earth as a neutrino of flavor \(\alpha\) with probability \(P_{\beta\alpha}(E, L_{S} - \ell)\).
Lastly, we have to sum or integrate over all unmeasured quantities, such as the energy of the primaries, the distance \(\ell\), all occuring particle species \(\eta\) form the source and the initially produced neutrino flavors \(\beta\).
As shown in Appendix~\ref{app:flux_lim} Eq.~\eqref{eq:flux_gen} contains Eq.~\eqref{eq:flux_simp} as a limiting case.\\
Now, we can turn towards the influence of quantum gravity effects on the flux spectra.
In the following, we consider two sensitive observables:
\begin{itemize}
	\item The total neutrino flux spectrum
	\item Neutrino flavor ratios
\end{itemize}
In order to discuss these observables, we introduce the threshold energy
\begin{align}
    E_{\mathrm{dip}} := \sqrt[\alpha]{\frac{M_{\mathrm{Planck}}^{\alpha-1}}{\min_{j}\vert d_{j} \vert L}} \,,
\end{align}
where quantum gravitational effects become relevant, i.e. where \(\Gamma_j L = \mathcal{O}(1)\).
From there on the oscillation probability approaches a uniform flavor distribution over all neutral fermions and the total probability for measuring any type of neutrino behaves as
\begin{align}
    \sum_{\beta = e}^{\tau} P_{\alpha\beta}(E \gg E_{\mathrm{dip}}) \rightarrow \frac{3}{N}\,,
\end{align}
i.e. it exhibits a dip after this threshold.
This is the crucial observation for everything we discuss from now on.


\subsubsection{Total Neutrino Fluxes}
\label{sssec:total_flux}
\begin{figure}
    \begin{subfigure}[b]{0.48\textwidth}
		\centering
		\includegraphics[width=\textwidth]{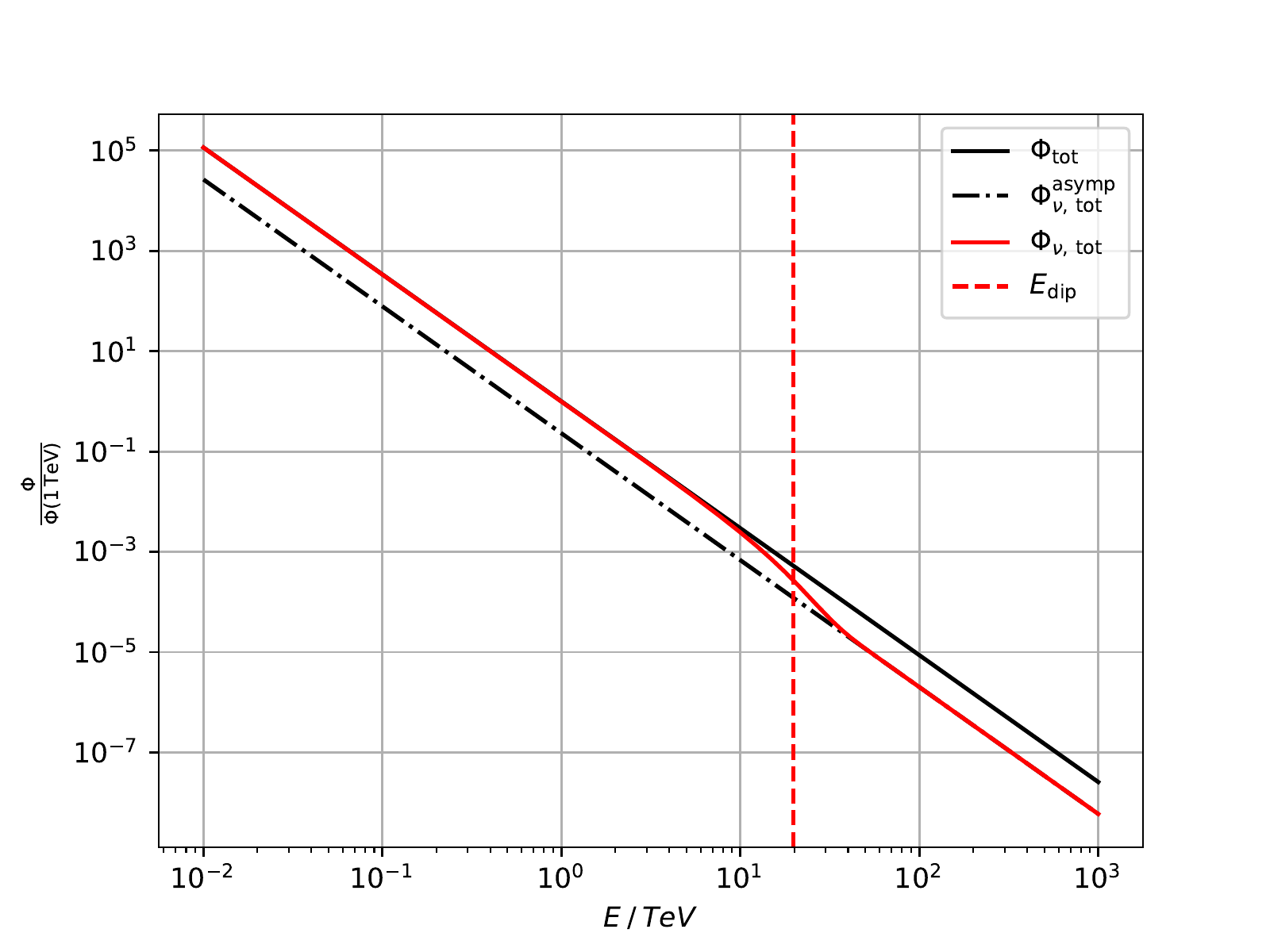}
 		\caption{}
 		\label{fig:flux_simp}
    \end{subfigure}
    \hfill
    \begin{subfigure}[b]{0.48\textwidth}
		\centering
		\includegraphics[width=\textwidth]{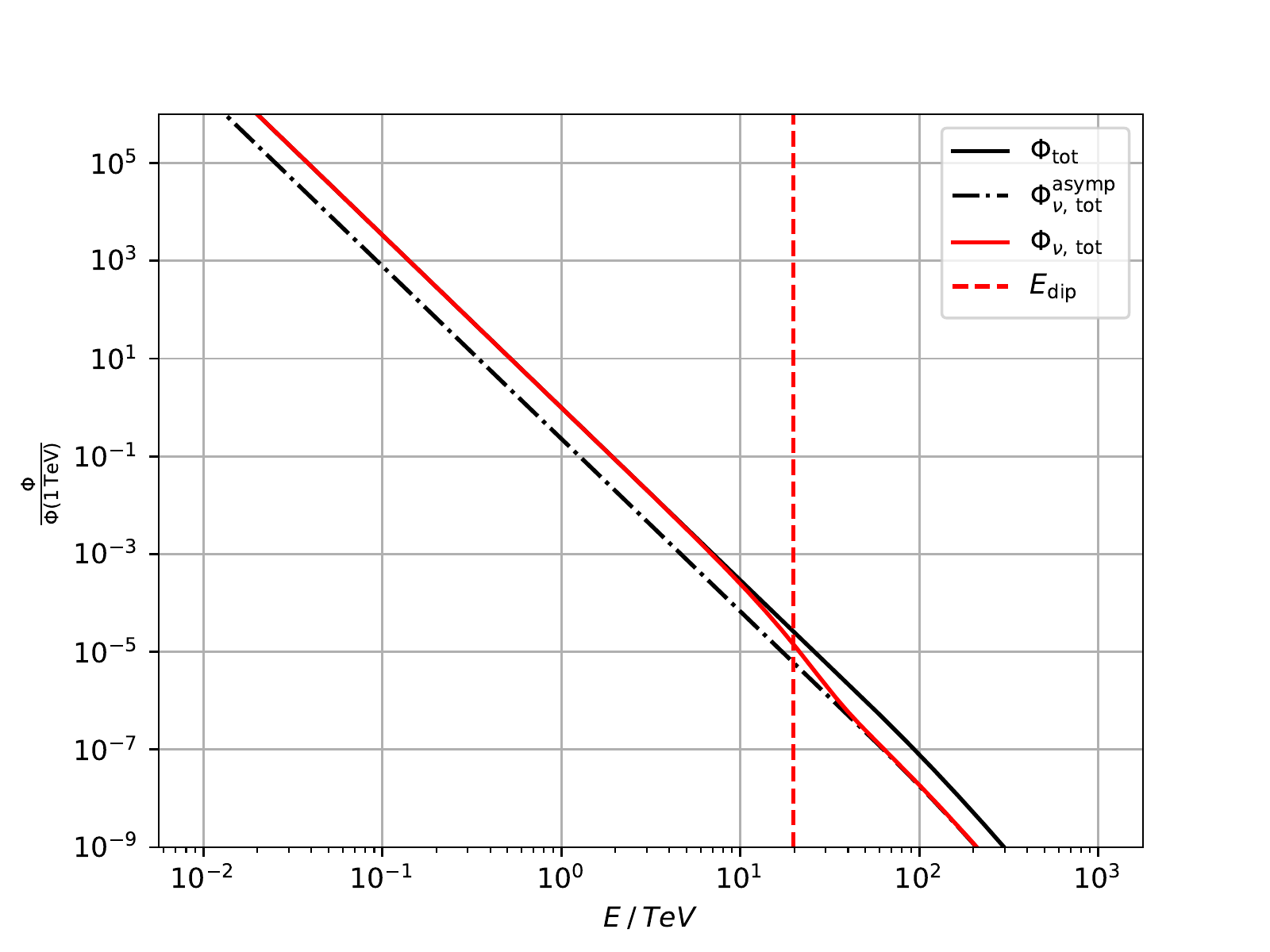}
  		\caption{}
  		\label{fig:flux_gen_app}
	\end{subfigure}
	\caption{Total Neutrino Flux \(\Phi_{\nu,\,\mathrm{tot}} = \sum_{\alpha}\Phi_{\alpha}^{\oplus}(E)\) for a primary (left) and secondary (right) neutrino source according to Eqs.~\eqref{eq:flux_simp} and~\eqref{eq:flux_gen_app}.
	The black line corresponds to the total particle flux reaching earth, while the red line represents \(\Phi_{\nu,\,\mathrm{tot}}\) calculated incluing decoherence effects.
	The black dashed line shows the asymptotic limit of \(\Phi_{\nu,\,\mathrm{tot}} = \sfrac{3}{N} \cdot \Phi_{\mathrm{tot}}\).
	Furthermore, we include a red, vertical, dashed line to show the position of the dip in the neutrino flux spectrum.}
    \label{fig:flux}
\end{figure}

According to the asymptotic behavior of the oscillation probabilities discussed in the last section, we expect a dip in the total neutrino flux spectrum beginning at the energy \(E_{\mathrm{dip}}\).
The sharpness of the dip will be influenced by whether we observe neutrinos originating from primary sources or from secondary ones and by the background of neutrinos from other sources.
This is because two neutrinos from the same secondary source will in general travel different distances depending on the point where their mother particles decay.
This shifts \(E_{\mathrm{dip}}\) to higher or lower values depending on the respective distance and hence the dip appears smeared out.
The same argument holds for two different neutrino sources at distances \(L_1\) and \(L_2\).\\
In Fig.~\ref{fig:flux_simp}, we plot the neutrino flux from a primary electron neutrino source with source flux~\cite{IceCube:2015rro, IceCube:2020acn}
\begin{align}
	\Phi_{e}^{S}(E) &= \Phi_0 E^{-\gamma}\,,\quad\mathrm{with}\;\gamma = 2.5\,.
\end{align}
The dip starts around \(E_{\mathrm{dip}}\) as expected and \(\Phi_{\nu,\,\mathrm{tot}}(E) := \sum_{\alpha}\Phi_{\alpha}^{\oplus}(E)\) quickly approaches the expected fraction of \(\sfrac{3}{N}\) compared to the initial flux.
The figure is obtained for \(L_{S} = 2\,\mathrm{kpc}\) and using the parameters from Tab.~\ref{tab:params}.
Here, we also choose \(L_S\) to be the approximate distance to the potentially interesting astrophysical neutrino source Cygnus OB2~\cite{Anchordoqui:2005gj}, as before.\\
As an example for a neutrino flux from a secondary source, we consider a source emmitting neutrons subsequently decaying into \(\bar{\nu}_e\).
For simplicity, we assume that neutrinos emerge from \(\beta\)-decay with a fixed mean energy \(\epsilon_0 \approx 0.5\,\mathrm{MeV}\) in the neutron rest frame.
Using this approximation and the parameters from Tab.~\ref{tab:params}, we can show that by summing over all final state neutrino flavors Eq.~\eqref{eq:flux_gen} becomes
\begin{align}
	\Phi_{\mathrm{tot}}(E) = &\frac{m_n}{2 \epsilon_0} \int\limits_{\frac{E m_n}{2\epsilon_0}}^{\infty}\,\diff E_n \frac{\Phi_{n}^{S}(E_n)}{E_n}\nonumber\\
	&\times \left[\frac{3}{N}\left(1 - e^{-\frac{L_{S}}{v_n \tau_n}}\right) + \frac{(N-3)}{N - N v_n \tau_n \Gamma}\left(e^{-\Gamma L_{S}} - e^{-\frac{L_{S}}{v_n \tau_n}}\right)\right]\,.
	\label{eq:flux_gen_app}
\end{align}
Here \(m_n\) is the neutron's mass, \(v_n\) represents its velocity and \(\tau_n\) is its mean lifetime in the lab frame.
The total neutrino fluxes obtained from Eq.~\eqref{eq:flux_gen_app} for the decoherence and standard cases can be seen in Fig.~\ref{fig:flux_gen_app}.
As expected the dip towards the asymptotic value \(\sfrac{3}{N} \cdot \Phi_{\mathrm{tot}}^{\mathrm{std}}\) is washed out compared to that from primary neutrino sources, but still occurs around \(E_{\mathrm{dip}}\).
Furthermore, the neutrino spectrum falls off more rapidly at the end of the considered energy range since at these energies the neutron mean free path amounts to
\begin{align*}
	\ell_n := v_n \tau_n = \frac{p_n}{E_n} \gamma \tau_{n}^{0}
	= \left. \frac{p_n}{m_n} \tau_{n}^{0} \: \right\vert_{E_n \approx 10\,\mathrm{PeV}} \approx 100\,\mathrm{pc}\,.
\end{align*}
Therefore, a few neutrons might even reach earth before decaying into neutrinos.\\
At this point, we can draw an intermediate conclusion.
Regardless of the kind of the source (primary or secondary) the total neutrino flux exhibits a characteristic dip if quantum gravity affects neutrino oscillation over astrophysical distances the way we have described it above.
The strength of this dip depends on the number of additional fermions \(N\) present in the beam, whereas its position and steepness depend on the model parameters \(d_{j}\) and \(\alpha\).
Hence observing such a dip in a neutrino flux spectrum immediately yields an upper bound on the number of neutral fermions in the universe and an estimate of the relevant decoherence parameters.

\subsubsection{Flavor Ratios at Neutrino Telescopes}
\label{sssec:flavor_ratios}
\begin{figure}
	\centering
	\includegraphics[width = 0.5\textwidth]{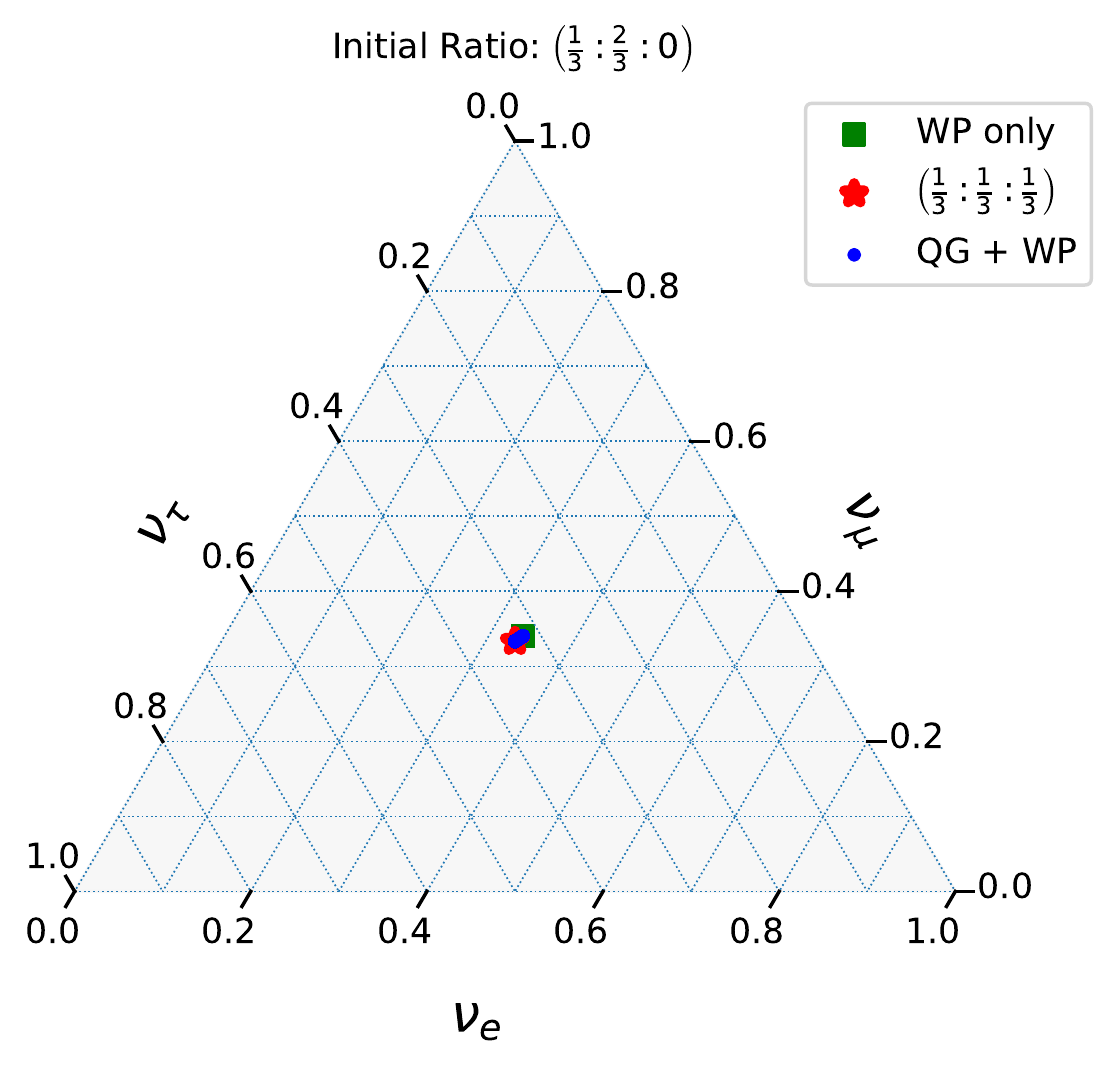}
	\caption{Final flavor ratios for \(E \in [1 \, \mathrm{TeV}, 1 \, \mathrm{PeV}]\) at \(L_{S} = 2 \, \mathrm{kpc}\) with \(r_{S} = (\sfrac{1}{3}:\sfrac{2}{3}:0)\) (pion source). The blue dots represent the ratios obtained from the full quantum gravity model, the green square shows the outcome of the wave-packet-decoherence-only case and the red star denotes full flavor equilibrium.}
	\label{fig:r_flav_pion}
\end{figure}

\begin{figure}
    \begin{subfigure}[b]{0.45\textwidth}
		\centering
		\includegraphics[width = \textwidth]{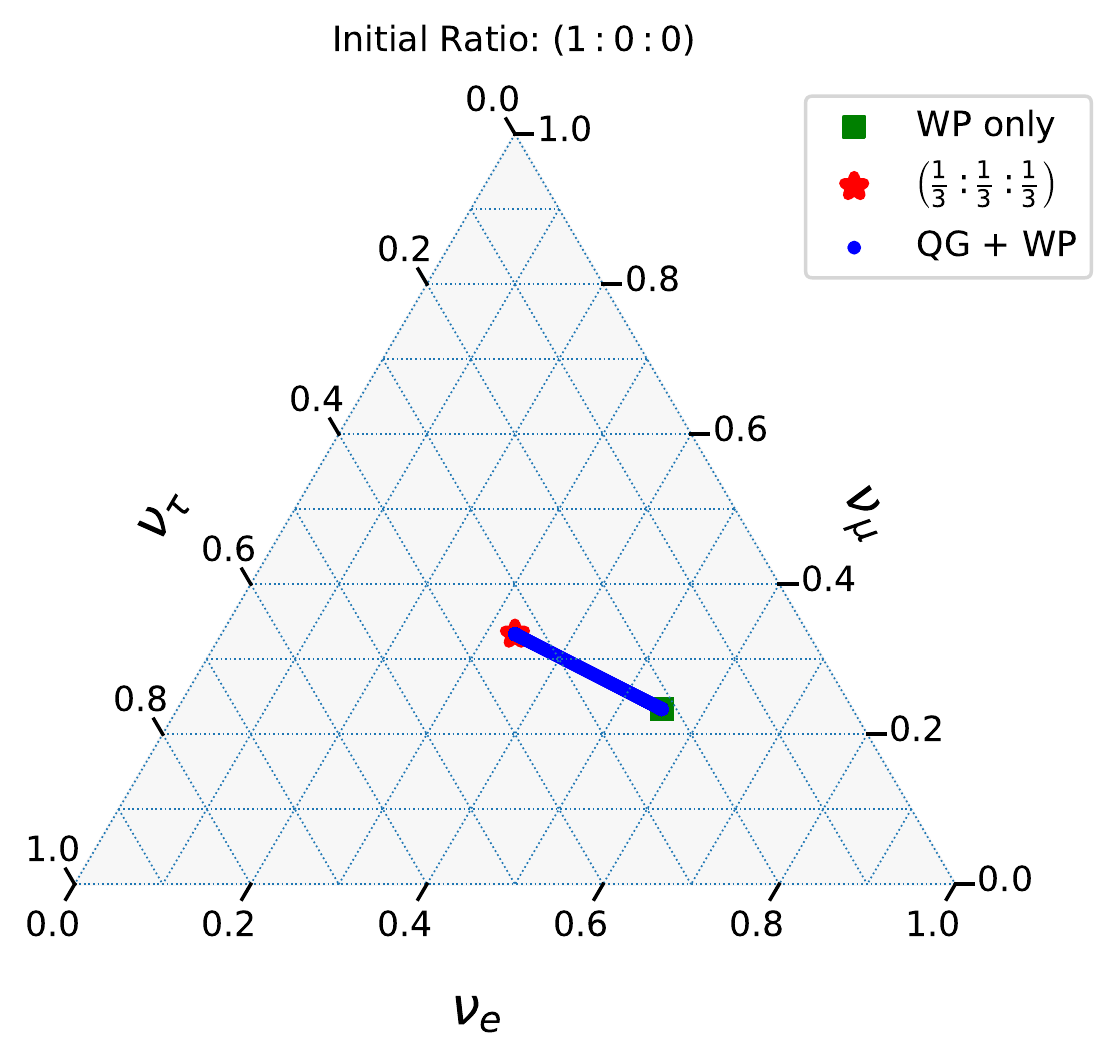}
		\caption{}
		\label{fig:r_flav_neutron1}
    \end{subfigure}
    \hfill
    \begin{subfigure}[b]{0.45\textwidth}
		\centering
		\includegraphics[width = \textwidth]{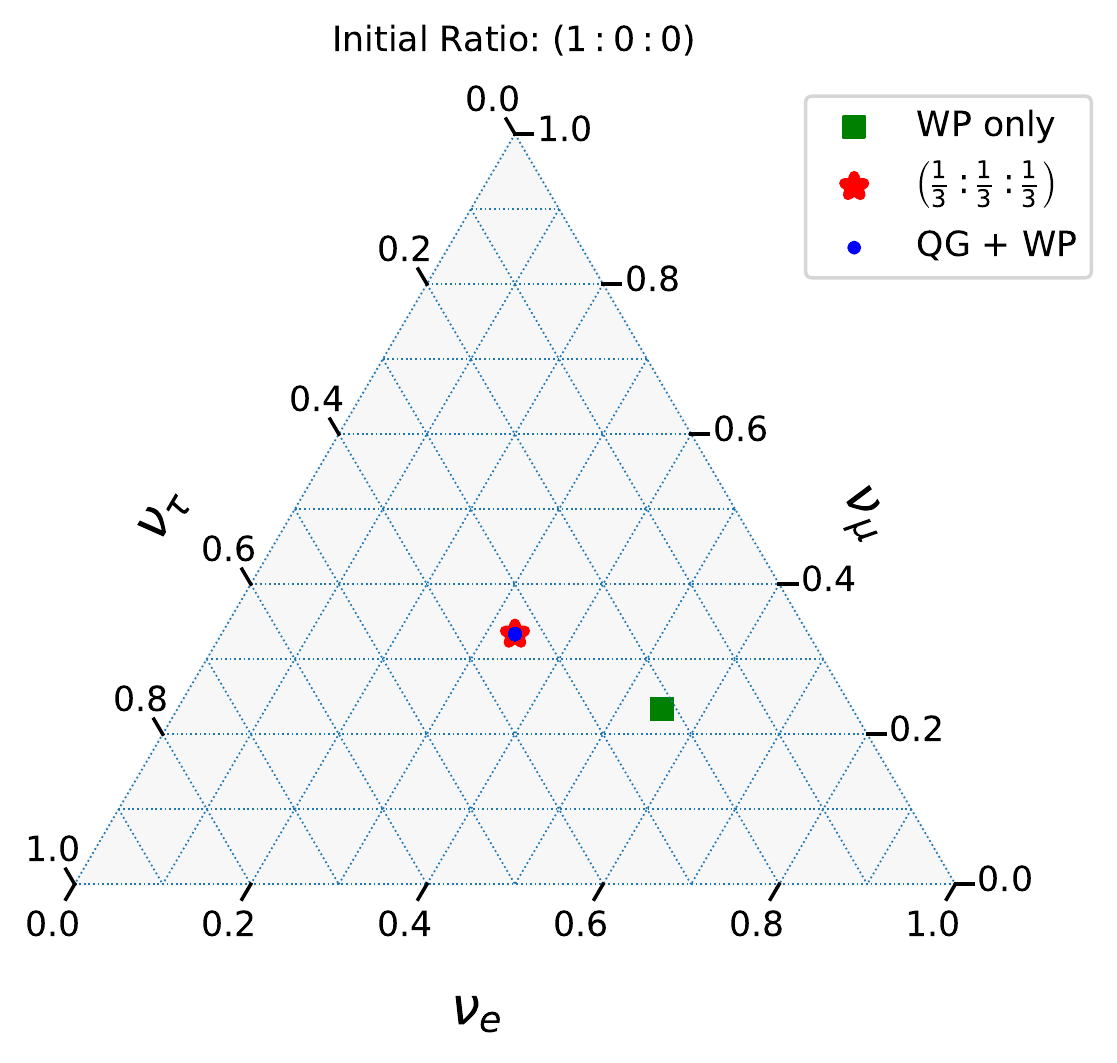}
		\caption{}
		\label{fig:r_flav_neutron2}
	\end{subfigure}
    \caption{Final flavor ratios for the respective energy intervals (\(E \in [1\,\mathrm{TeV}, 1\,\mathrm{PeV}]\) left and \(E \in [3 E_{\mathrm{dip}}, 1\,\mathrm{PeV}]\) right) at \(L_{S} = 2 \, \mathrm{kpc}\) with \(r_{S} = (1:0:0)\) (neutron source). The blue dots represent the ratios obtained from the full quantum gravity model, the green square shows the outcome of the wave-packet-decoherence-only case and the red star denotes full flavor equilibrium.}
    \label{fig:r_flav_neutron}
\end{figure}

\begin{figure}
    \begin{subfigure}[b]{0.45\textwidth}
		\centering
		\includegraphics[width = \textwidth]{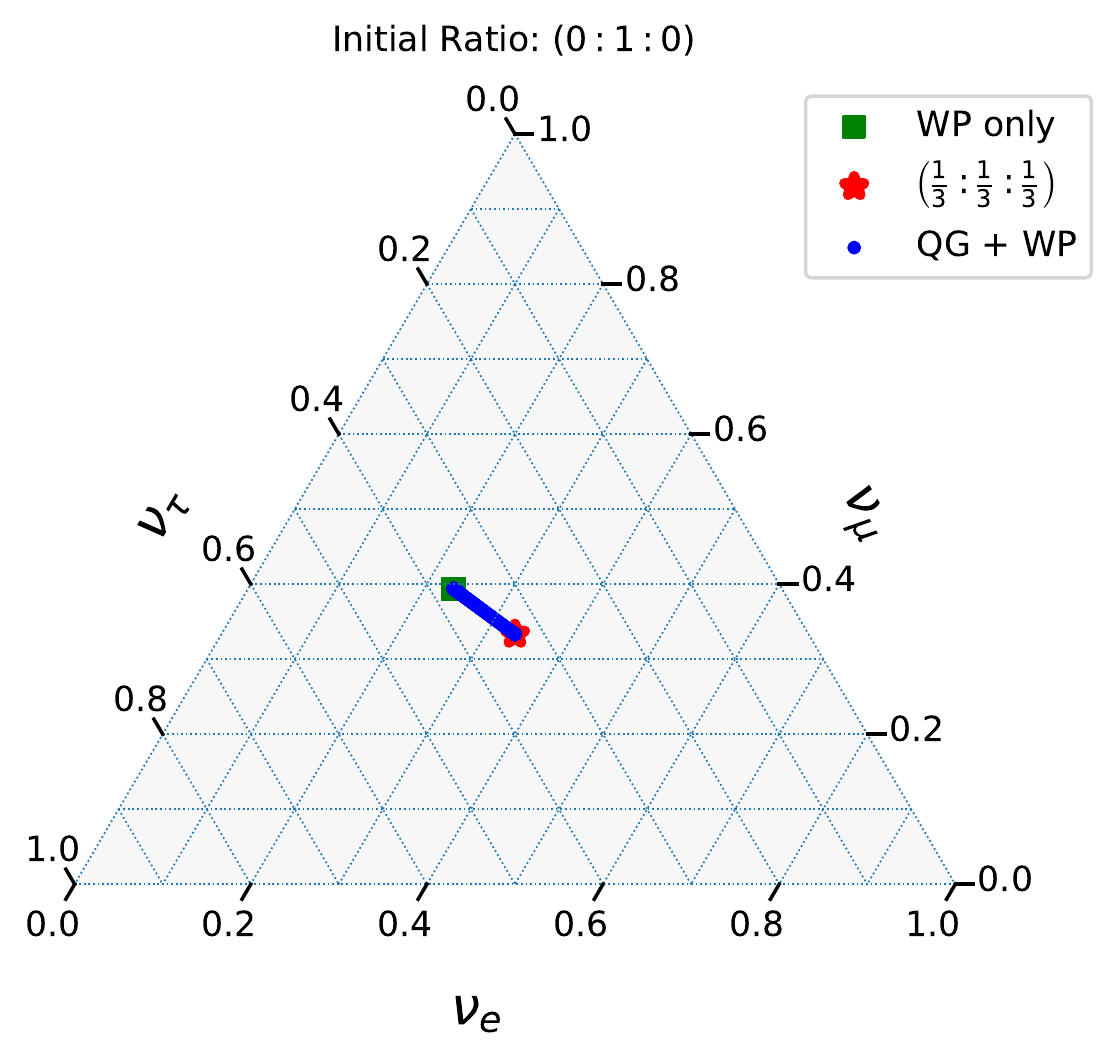}
		\caption{}
		\label{fig:r_flav_mudamp1}
    \end{subfigure}
    \hfill
    \begin{subfigure}[b]{0.45\textwidth}
		\centering
		\includegraphics[width = \textwidth]{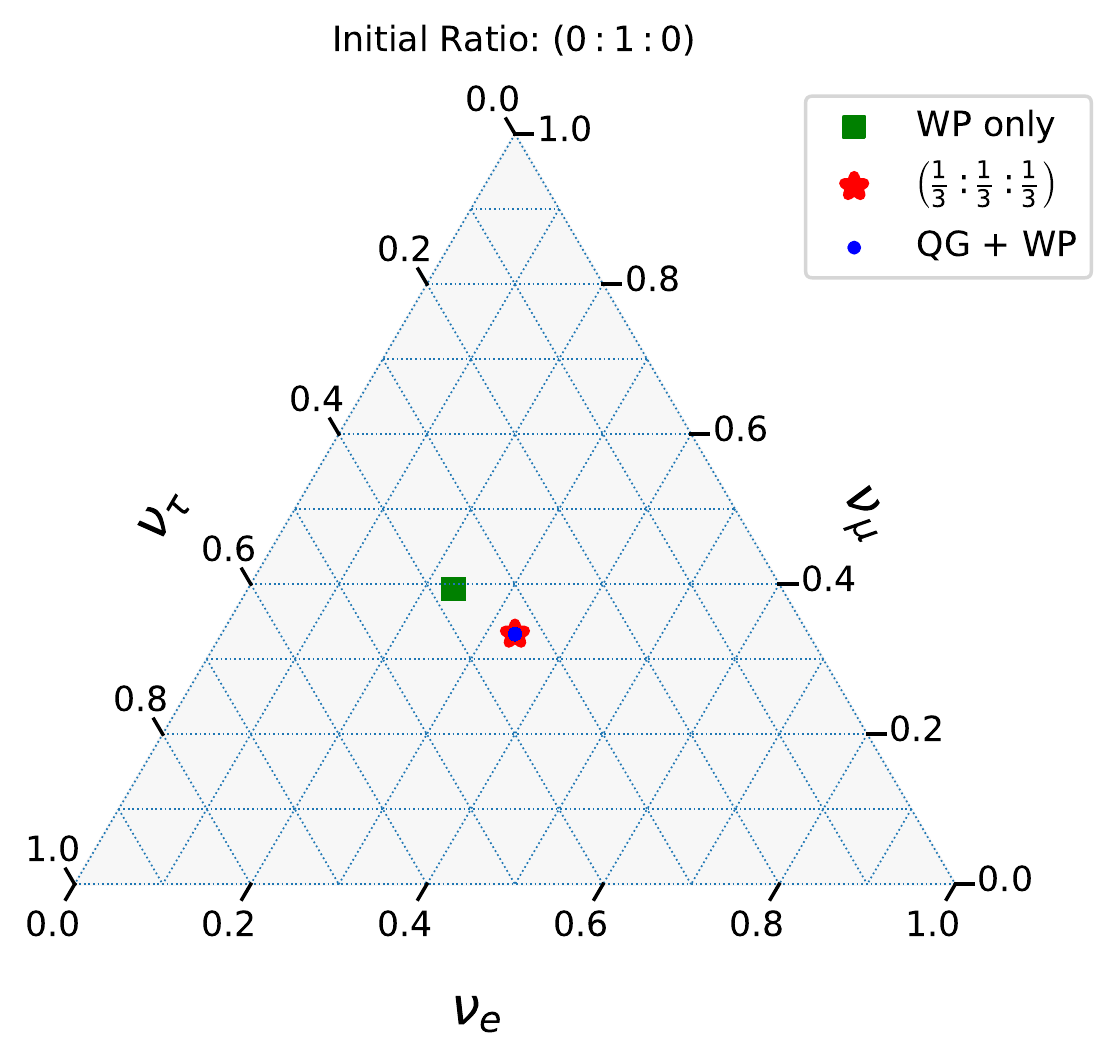}
		\caption{}
		\label{fig:r_flav_mudamp2}
	\end{subfigure}
    \caption{Final flavor ratios for the respective energy intervals (\(E \in [1\,\mathrm{TeV}, 1\,\mathrm{PeV}]\) left and \(E \in [3 E_{\mathrm{dip}}, 1\,\mathrm{PeV}]\) right) at \(L_{S} = 2 \, \mathrm{kpc}\) with \(r_{S} = (0:1:0)\) (muon damped source). The blue dots represent the ratios obtained from the full quantum gravity model, the green square shows the outcome of the wave-packet-decoherence-only case and the red star denotes full flavor equilibrium.}
    \label{fig:r_flav_mudamp}
\end{figure}

The other QG sensitive observables at neutrino telescopes are the reconstructed neutrino flux ratios~\cite{Anchordoqui:2005gj}.
They depend on the flavor composition \(r_S\) at the source and on the details of the evolution of the system.
Especially, we expect flavor equilibrium, i.e.
\begin{align}
    \Phi_{e}(E) \simeq \Phi_{\mu}(E) \simeq \Phi_{\tau}(E)\,,
\end{align}
for energies where QG effects become relevant, i.e. at \(E \geq E_{\mathrm{dip}}\), according to the asymptotic behavior of the oscillation probabilities.\\
In the following, we denote the neutrino flavor ratios as
\begin{align}
	r &= (r^{e} : r^{\mu} : r^{\tau}) \\
	&= \left( \frac{\Phi_{e}}{\Phi_{\nu,\mathrm{tot}}} :
	\frac{\Phi_{\mu}}{\Phi_{\nu,\mathrm{tot}}} :
	\frac{\Phi_{\tau}}{\Phi_{\nu,\mathrm{tot}}} \right)\,,
\end{align}
such that a pion source producing two \(\nu_\mu\) per each \(\nu_e\) and no \(\nu_\tau\) yields an initial flavor ratio of
\begin{align}
	r_{S}^{\pi^{\pm}} &= \left(\frac{1}{3} : \frac{2}{3} : 0\right)\,.
\end{align}
Although flavor ratios are insensitive to the number of additional fermions in our model, they still provide insight about if quantum gravity effects are present.
In the case of democratic quantum gravity effects, flavor ratios always approach a \((\sfrac{1}{3}:\sfrac{1}{3}:\sfrac{1}{3})\) ratio at high energies regardless of the initial flavor composition.\\
For an initial pion source, as exemplified above, it is important to note that this signature is already expected for wave packet decoherence effects due to the maximal mixing of \(\nu_\mu\) and \(\nu_\tau\)~\cite{Pakvasa:2007dc,Bustamante:2015waa}.
Hence, it would be beneficial to examine multiple sources of different initial flavor compositions in order to tell both effects from each other.
In the following, we mainly focus on three kinds of idealized sources as they are the most commonly used ones~\cite{Anchordoqui:2005gj,Pakvasa:2007dc}
\begin{itemize}
	\item Pion source \(\Leftrightarrow\) \(r_S = (\sfrac{1}{3}:\sfrac{2}{3}:0)\)
	\item Neutron source \(\Leftrightarrow\) \(r_S = (1:0:0)\)
	\item Muon damped pion source \(\Leftrightarrow\) \(r_S = (0:1:0)\)
\end{itemize}
This is of course not an exhaustive list and it was shown~\cite{Pakvasa:2007dc} that one has to be careful with assuming such idealized scenarios in order to infer neutrino oscillation parameters from experiment.
In our case this does not play a role since we only want to demonstrate how the impact of quantum gravitational decoherence alters the observed flavor ratios at the detector.
To do so, we compare for each example the respective final flavor compositions of wave packet separation decoherence only and with additional quantum gravitational decoherence.\vspace{0.2cm}\\
\textbf{Pion Source:}\\
Fig.~\ref{fig:r_flav_pion} shows the final flavor ratios for a pion source at different energies.
Here, we can see that for an initial flavor ratio of \((\sfrac{1}{3}:\sfrac{2}{3}: 0)\) it is difficult to distinguish pure wave packet decoherence from additional quantum gravitational decoherence effects, since both lead to a final uniform flavor distribution, as discussed above.
For arbitrary high statistics and very low systematical error it might be possible for experiments to tell both cases apart, but such benefitial conditions are only to be expected in the very far future, if at all.\vspace{0.2cm}\\
\textbf{Neutron Source:}\\
For an astrophysical neutron source giving rise to an initial ratio of \((1:0:0)\), we obtain a different picture as can be inferred from Figs.~\ref{fig:r_flav_neutron1} and~\ref{fig:r_flav_neutron2}.
Both plots show the final flavor ratios for different energy intervals \(E \in [1\,\mathrm{TeV}, 1\,\mathrm{PeV}]\) and \(E \in [3 E_{\mathrm{dip}}, 1\,\mathrm{PeV}]\), respectively.
The first interval includes energies below the critical energy \(E_{\mathrm{dip}} \sim 20 \, \mathrm{TeV}\) which is why in Fig.~\ref{fig:r_flav_neutron1} the QG decoherence points start at the WP only point and approach the democratic scenario for increasing energy.
The second interval is chosen to show that after the threshold \(E_{\mathrm{dip}}\) all flavor ratios are located at the \((\sfrac{1}{3}:\sfrac{1}{3}:\sfrac{1}{3})\) point.\\
In both Figs., pure wave packet decoherence and quantum gravitational decoherence can be nicely distinguished as soon as the confidence intervals of the experimental measurement is small enough which requires significantly less statistics compared to the former case.\vspace{0.2cm}\\
\textbf{Muon damped pion Source:}\\
For a muon damped pion source we assume that the muon from the pion decay quickly looses energy to the surrounding matter.
Consequently, the \(\nu_\mu\) and \(\nu_e\) following from the decaying muon also have much less energy than the \(\nu_\mu\) released during pion decay.
Hence, even if they reach the detector they would not be regarded as high energy neutrinos and discarded.
In this case, the effective initia flavor ratio is \((0:1:0)\) and the resulting final ratios are shown in Figs.~\ref{fig:r_flav_mudamp1} and~\ref{fig:r_flav_mudamp2}.
As for the neutron source, we show plots for both energy intervalls.

\section{Conclusions}
\label{sec:conc}
Quantum gravitational (QG) decoherence in astrophysical neutrinos could provide important insights into the nature and composition of dark matter.
A thorough analysis using the formalism of neutrino density matrices shows that a wide class of possible decoherence models leads to a uniform flavor distribution over all fermionic degrees of freedom.
These additional fermions need to carry the same unbroken gauge quantum numbers as neutrinos in order to be indistinguishable by interactions with the spacetime foam.
Hence, all fermions contributing to dark matter with masses smaller than the beam energy can in principle be observed using this effect.
The impact of quantum gravity decoherence might only become visible for highly energetic neutrinos which traveled very long distances since these effects might be very weak for non-Planck scale neutrinos.\\
If a dip similar to the ones shown in Figs.~\ref{fig:flux_simp} and~\ref{fig:flux_gen_app} is present in the total measured neutrino flux, this might be a hint at the quantum properties of spacetime in combination with the presence of several dark matter fermions.
Moreover, the depth of the dip is a measure for how many of these fermionic species exist.\\
Even if we don't see a dip as shown in Sec.~\ref{ssec:flux} quantum gravitational decoherence enforces a uniform flavor distribution over all neutrino species.
Thus, one observes a \((\sfrac{1}{3}:\sfrac{1}{3}:\sfrac{1}{3})\) flavor ratio in the high energy region of astrophysical neutrino fluxes regardless of the flavor composition at the source.
Depending on the original flavor composition at the source, already this democratic flavor mix may hint at the quantum properties of spacetime and that either interactions with the spacetime foam do not obey the no-hair theorem or that dark matter is not made up from fermions with masses smaller than the beam energy.\\
Finally because of the inherent energy dependence of the neutrino flavor ratios, QG implies that the usual ansatz parametrizing the neutrino fluxes as
\begin{align}
    \Phi_{\alpha} &= f_{\alpha} C E^{-\gamma} \,,
\end{align}
where the flavor fraction \(f_{\alpha}\) is constant with respect to energy, is no longer valid.
According to our previous findings regarding the total neutrino flux as well as neutrino flavor ratios the flavor composition at the detector can actually vary significantly with respect to energy.\\
In conclusion, we see that the observation of highly energetic neutrinos of astrophysical origin at future experiments bears significant potential regarding the search for effects beyond the Standard Model of particle physics.
Hence, pushing efforts to observe a significant flux of these neutrinos should be considered.

\section*{Acknowledgements}
ER is supported by The 5000 Doktor scholarship Program by The Ministry of Religious Affairs, Indonesia.

\printbibliography

\newpage
\appendix
\section{Classification of dissipators leading to flavor equilibrium}
\label{app:flav_eq}
Here, we want to show briefly which dissipators will inevitably lead to an asymptotic final state of maximal entropy, considering an \(N\) level system as we do in this paper.
For the case of neutrino oscillations this corresponds to maximal flavor equilibrium.
A maximally entropic state is described by a density matrix proportional to the identity or, if we use the language of the basis matrices from Sec.~\ref{sec:model}, which is proportional to \(\lambda_0\), i.e.
\begin{align}
    \rho &= \varrho_0 \lambda_0 \,.
\end{align}
This is achieved by any dissipator \(\mathcal{D}\) damping all other componentes but \(\varrho_0\) in the asymptotic limit.
Recall that the Lindblad equation reads
\begin{align}
    \frac{\mathrm{d}}{\mathrm{d} x} \rho(x) &= -i[H,\rho(x)] + \mathcal{D}[\rho] \,.
\end{align}
By choosing an operator basis in which we can expand the density matrix and the hamiltonian as
\begin{align}
    \rho &= \varrho_0 \lambda_0 + \vec{\varrho} \cdot \vec{\lambda} \,, \\
    H &= h_0 \lambda_0 + \vec{h} \cdot \vec{\lambda} \,,
\end{align}
we can rewrite this for a trace preserving system as
\begin{align}
    \dot{\varrho}_0 &= 0 \,, \\
    \dot{\vec{\varrho}} &= \tilde{C} \vec{\varrho} + \tilde{D} \vec{\varrho} \,.
\end{align}
Here \(\tilde{C} = -\tilde{C}^T\) and \(\tilde{D}\) are the representation matrices of the commutator part and the dissipator, respectively, reduced by the zeroth row and column.
The commutator part of this equation conserves the length of \(\varrho\) while a nontrivial dissipator will change it, since
\begin{align}
    \frac{\mathrm{d}}{\mathrm{d}t} (\vec{\varrho}^{\,T} \cdot \vec{\varrho}) &= \dot{\vec{\varrho}}^{\,T} \cdot \vec{\varrho} +  \vec{\varrho}^{\,T} \dot{\vec{\varrho}} \\
    &= (\tilde{C} \vec{\varrho} + \tilde{D} \vec{\varrho})^T \vec{\varrho} +  \vec{\varrho}^{\,T} (\tilde{C} \vec{\varrho} + \tilde{D} \vec{\varrho}) \\
    &= \vec{\varrho}^{\,T}\tilde{C}^T \vec{\varrho} + \vec{\varrho}^{\,T}\tilde{C} \vec{\varrho} + \vec{\varrho}^{\,T}\tilde{D}^T \vec{\varrho} + \vec{\varrho}^{\,T}\tilde{D} \vec{\varrho} \\
    &= -\vec{\varrho}^{\,T}\tilde{C} \vec{\varrho} + \vec{\varrho}^{\,T}\tilde{C} \vec{\varrho} + \vec{\varrho}^{\,T}\tilde{D}^T \vec{\varrho} + \vec{\varrho}^{\,T}\tilde{D} \vec{\varrho}\\
    &= \vec{\varrho}^{\,T}\tilde{D}^T \vec{\varrho} + \vec{\varrho}^{\,T}\tilde{D} \vec{\varrho} \,.
\end{align}
Furthermore, we assume the dissipator to be a symmetric matrix, i.e. \(\tilde{D} = \tilde{D}^T\) (which is usually the case) and as such it can be diagonalized using an orthogonal matrix \(O \in O(N^2-1)\), i.e.
\begin{align}
     \tilde{D} &= O \tilde{\Delta} O^T\,,
\end{align}
where \(\tilde{\Delta} = \mathrm{diag}(\delta_1, \ldots, \delta_{N^2-1})\) contains the eigenvalues of \(\tilde{D}\) on its diagonal.\\
This orthogonal transformation corresponds to a partial change of basis where only \(\lambda_0\) remains unchanged.
Applying this procedure to the Lindblad equation yields
\begin{align}
    \dot{\vec{\varrho}} &= O \tilde{C}^{\prime} O^T \vec{\varrho} + O \tilde{\Delta} O^T \vec{\varrho} \\
    O^T \dot{\vec{\varrho}} &= \tilde{C}^{\prime} O^T \vec{\varrho} + \tilde{\Delta} O^T \vec{\varrho} \\
    \dot{\vec{\varrho}}^{\,\prime} &= \tilde{C}^{\prime}\vec{\varrho}^{\,\prime} + \tilde{\Delta}\vec{\varrho}^{\,\prime} \,.
\end{align}
Here we introduce the transformed quantities \(\vec{\varrho}^{\,\prime} = O^T \vec{\varrho}\) and \(\tilde{C}^\prime = O^T \tilde{C} O\) and moreover assume that \(\dot{O} \equiv 0\).
Since the commutator part remains antisymmetric under this transformation,
\begin{align}
    (\tilde{C}^\prime)^T = (O^T \tilde{C} O)^T = O^T \tilde{C}^T O = - O^T \tilde{C} O = -\tilde{C}^\prime \,,
\end{align}
it still preserves the length of \(\vec{\varrho}\) as shown above.\\
Hence, the square of \(\vec{\varrho}\) evolves according to
\begin{align}
    \frac{\mathrm{d}}{\mathrm{d} t} \vec{\varrho}^{\,T} \vec{\varrho} &= 2 \vec{\varrho}^{\,T} \tilde{D} \vec{\varrho} \\
    &= 2 \vec{\varrho}^{\,T} O O^T  \tilde{D} O O^T \vec{\varrho} \\
    &= 2 \vec{\varrho}^{\,\prime \, T} O^T \tilde{D} O \vec{\varrho}^{\,\prime} \\
    &= 2 \vec{\varrho}^{\,\prime \, T} \tilde{\Delta} \vec{\varrho}^{\,\prime} \\
    &= 2 \sum_{k = 1}^{N^2 - 1} \delta_{k} (\varrho^{\,\prime}_{k})^2 \,.
\end{align}
In case \(\delta_k < 0\) for all \(k \geq 1\), the system is asymptotically damped to the identity, i.e. \(\rho = \varrho_0 \lambda_0\), regardless of the initial state of the system.
Consequently all negatively definit dissipators lead to a maximally entropic final state in the asymptotic limit.

\section{Instantaneous decay approximation of the flux formula}
\label{app:flux_lim}
Now, we verify that Eq.~\eqref{eq:flux_gen} reduces to Eq.~\eqref{eq:flux_simp} if the emerging \(\eta\) particles decay instantaneously into neutrinos of flavor \(\beta\) with the same energy, i.e. \(S\) is a source emmitting only neutrinos.
This means \(\eta\) represents a mere mathematical tool to relate both formulas.
This corresponds to
\begin{align}
	\pi_{\eta\beta}(E, E_{\eta}) &= \delta(E - E_{\eta}) \delta_{\eta\beta}\,,\\
	\tau_{\eta} &\rightarrow 0^+\,,
\end{align}
and yields
\begin{align}
	\Phi_{\alpha}^{\oplus}(E) &= \sum_{\eta \in S}\sum_{\beta = e}^{\tau} \; \int\limits_{m_{\eta}}^{\infty} \Phi_{\eta}^{S}(E_{\eta})
	\pi_{\eta \beta}(E, E_{\eta}) \lim_{\tau_{\eta} \rightarrow 0^+}\int\limits_{0}^{L_{S}} \, \frac{e^{\frac{-\ell}{v_{\eta}\tau_{\eta}}}}{v_{\eta}\tau_{\eta}}
	P_{\beta \alpha}(E, L_{S} - \ell) \,\diff\ell \diff E_{\eta}\\
	&= \sum_{\eta \in S}\sum_{\beta = e}^{\tau} \; \int\limits_{m_{\eta}}^{\infty} \Phi_{\eta}^{S}(E_{\eta})
	\delta(E - E_{\eta}) \delta_{\eta\beta} \lim_{\tau_{\eta} \rightarrow 0^+}\int\limits_{0}^{L_{S}} \, \frac{e^{\frac{-\ell}{v_{\eta}\tau_{\eta}}}}{v_{\eta}\tau_{\eta}}
	P_{\beta \alpha}(E, L_{S} - \ell) \,\diff\ell \diff E_{\eta}\\
	&= \sum_{\beta = e}^{\tau} \Phi_{\beta}^{S}(E) \lim_{\tau_{\eta} \rightarrow 0^+}\int\limits_{0}^{L_{S}} \, \frac{e^{\frac{-\ell}{v_{\eta}\tau_{\eta}}}}{v_{\eta}\tau_{\eta}}
	P_{\beta \alpha}(E, L_{S} - \ell) \,\diff\ell \diff E_{\eta}\\
	&= \sum_{\beta = e}^{\tau} P_{\beta \alpha}(E, L_{S}) \Phi_{\beta}^{S}(E)\,.
\end{align}
From the second step to the last step, we used the relation
\begin{align}
    I := \lim_{\tau \rightarrow 0^+} \int\limits_{0}^{L} \frac{e^{-\frac{\ell}{v \tau}}}{v \tau} P_{\beta\alpha}(E, L - \ell) \,\diff\ell = P_{\beta\alpha}(E, L)\,.
\end{align}
\textit{Proof:}\\
To derive this, we simplify the expression above by introducing \(T := \sfrac{L}{v}\), changing variables from \(\ell\) to \(t := \sfrac{\ell}{v}\) and defining \(g(t) := P_{\beta\alpha}(E, L - vt)\).
This yields
\begin{align}
    I = \lim_{\tau \rightarrow 0^+} \int\limits_{0}^{T} \frac{e^{-\frac{t}{\tau}}}{\tau} g(t) \,\diff t \stackrel{!}{=} g(0)\,.
\end{align}
In the following, we use the definition of \(\lim_{\tau \rightarrow 0^+}\) and choose an arbitrary but positive null sequence \((\tau_{n})_{n \in \mathbb{N}}\) replacing
\begin{align}
    \lim_{\tau \rightarrow 0^+} \rightarrow \lim_{n \rightarrow \infty}\,.
\end{align}
Therefore, we get
\begin{align}
    I &= \lim_{n \rightarrow \infty} \int\limits_{0}^{T} \underbrace{\frac{e^{-\frac{t}{\tau_{n}}}}{\tau_{n}}}_{\geq 0} g(t) \,\diff t \\
    &= \lim_{n \rightarrow \infty} g(\xi_n) \int\limits_{0}^{T} \frac{e^{-\frac{t}{\tau_{n}}}}{\tau_{n}} \,\diff t \\
    &= \lim_{n \rightarrow \infty} g(\xi_n) \left(1 - e^{-\frac{T}{\tau_n}}\right) \\
    &= g(0) \,.
\end{align}
This is what we had to show.\\
Here we made use of the generalized mean value theorem of integration and exploited that \(\xi_n \rightarrow 0\), since
\begin{align}
    g(t)e^{-\frac{t}{\tau}} \rightarrow 0
\end{align}
sufficiently rapid for all \(t > 0\), due to the exponential decay of the integrand.

\end{document}